    \documentclass[fleqn,usenatbib]{mnras}
    
    \usepackage{newtxtext,newtxmath}
    \usepackage{soul}
    
    \usepackage[T1]{fontenc}
    
    \DeclareRobustCommand{\VAN}[3]{#2}
    \let\VANthebibliography\thebibliography
    \def\thebibliography{\DeclareRobustCommand{\VAN}[3]{##3}\VANthebibliography}
    
    \usepackage{natbib}
    \usepackage{hyperref}
    \defcitealias{2020Winter}{WKC20}
    \defcitealias{2015Matt}{MBB15}
    
    \usepackage{graphicx}	
    \usepackage{amsmath}	
    \usepackage{soul}
    
    \title[Local-FUV radiation impact on spin evolution]{\textit{The influence of the environment on the spin evolution of low-mass stars. }\\I. External photoevaporation of circumstellar disks}
    
    \author[J. Roquette et al.]{
    J. Roquette,$^{1}$\thanks{E-mail: jt574@exeter.ac.uk}, S. P. Matt$^{1}$,  A.~J.~Winter$^{2}$, L. Amard$^{1}$,  and S. Stasevic$^{1}$
    \\
    $^{1}$University of Exeter, Physics and Astrophysics dept, Stoker Road, EX44QL, Exeter, UK\\
    $^{2}$ Institut f\"ur Theoretische Astrophysik, Zentrum f\"ur Astronomie, Heidelberg University, Albert Ueberle Str. 2, 69120 Heidelberg,
    Germany\\
    }
    
    \date{Accepted XXX. Received YYY; in original form ZZZ}
    
    \pubyear{2021}
    
\begin{document}

    \label{firstpage}
    \pagerange{\pageref{firstpage}--\pageref{lastpage}}
    \maketitle
    
    \begin{abstract}
    Massive stars are strong sources of far-ultraviolet radiation that can be hostile to the evolution of protoplanetary disks, driving mass loss by external photoevaporation and shortening disk-dissipation timescales. Their effect may also reduce the timescale of angular momentum exchanges between the disk and host star during the early pre-main-sequence phase. To improve our understanding of the environmental influence on the rotational history of stars, we developed a model that considers the influence of the local far-ultraviolet radiation on the spin evolution of low mass stars. Our model includes an assumption of disk-locking, which fixes the rotation rate during the star-disk-interaction phase, with the duration of this phase parametrised as a function of the local far-ultraviolet radiation and stellar mass (in the range 0.1--1.3 M$_\odot$). In this way, we demonstrate how the feedback from massive stars can significantly influence the spin evolution of stars and explain the mass-dependency observed in period-mass distributions of young regions like Upper Sco and NGC 2264. The high far-ultraviolet environments of high-mass stars can skew the period distribution of surrounding stars towards fast-rotation, explaining the excess of fast-rotating stars in the open cluster h Per. The proposed link between rotation and the pre-main-sequence environment opens new avenues for interpreting the rotational distributions of young stars. For example, we suggest that stellar rotation may be used as a tracer for the primordial ultraviolet irradiation for stars up to $\sim$1 Gyr, which offers a potential method to connect mature planetary systems to their birth environment. 
    
    \end{abstract}
    
    
    
    
    \section{Introduction}
    \label{sec:introduction}
    
    Over the last couple of decades, a growing number of studies have been exploring the influence of the environment on the evolution and dissipation of protoplanetary disks \citep[\emph{e.g.},][]{1998Bally,1998Johnstone,2004Adams,2010Adams,2016Faccini,2019Nicholson,2019HaworthClarke,2020Sellek,2021Parker}.  Of particular importance is the existence of massive stars in a cluster's population \citep{2004Adams,2018Winter}. These massive stars can radically influence their surrounding environment due to their strong far-ultraviolet (FUV) emission.  FUV photons (6eV$<h\nu<$ 13.6eV) can dissociate H$_2$ molecules and trigger the external photoevaporation of the material in protoplanetary disks around neighbouring low mass stars, shortening the timescales of disk evolution and dissipation \citep{2004Adams,2020Winter,2021ConchaRamirezExternalPhotoevaporation}. A number of studies have observationally explored the indirect consequences of external photoevaporation of disks, such as the study of proplyd objects in the neighbourhood of massive stars \citep{2003Smith,2006Balog,2009Rigliaco,2012Wright,2014Mann,2014Guarcello,2016Kim,2021Haworth} and variations of the disk-fraction and disk-masses of stars in clusters, depending on their proximity to massive stars \citep{2007Balog,2010Guarcello,2012Fang,2016Guarcello,2017Ansdell,2018Eisner}. Because of its importance on constraining the timescales for planet-formation \citep[\emph{e.g.},][]{2000Armitage,2014NajitaKenyon,2019Concha-Ramirez,2020Ansdell}, the dependence of the disk-dissipation timescales with the local FUV flux has been recently theoretically investigated by a number of authors \citep[\emph{e.g.},][]{2004Adams,2016Faccini,2018Haworth,2018Winter,2019ConcharRamirez_b,2020Winter,2020Sellek,2021Parker}. 
    
    \begin{figure*}
        \centering
        \includegraphics[width=1.5\columnwidth]{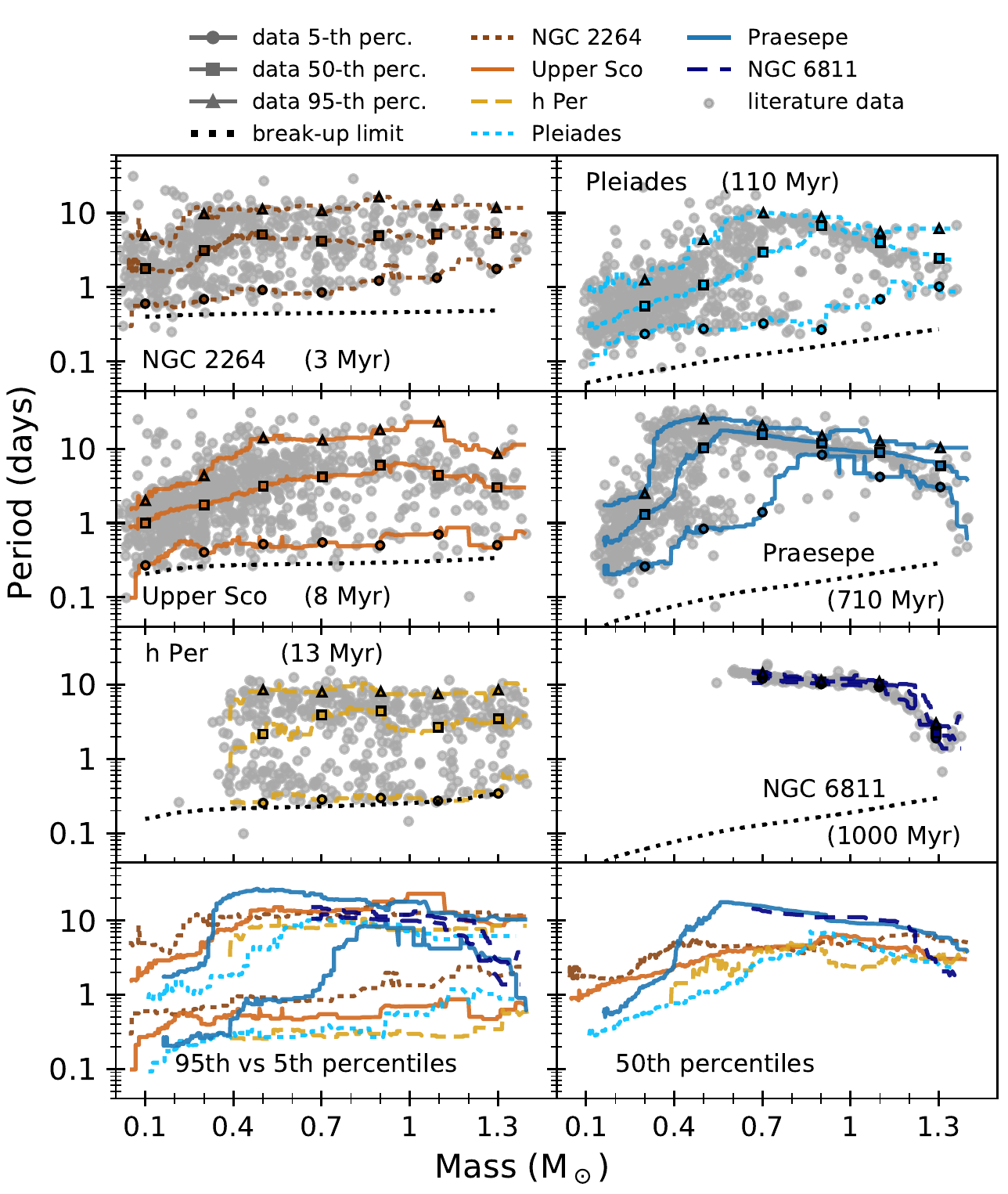}
        \caption{Observed rotational period distributions of low mass stars ($M_*\leq1.4\mathrm{M}_\odot$) in clusters and associations at 3 Myr (NGC 2264), 8 Myr (Upper Sco), 13 Myr (h Per), 110 Myr (Pleiades), 710 Myr (Praesepe) and 1000 Myr (NGC 6811). Observed rotational periods are plotted as a function of stellar mass. Panels are ordered as a function of age. The dashed lines show the break-up limit at each age. The dashed-symbol lines show the 5th (circle), 50th (square) and 95th (triangle) rolling percentiles of the period-mass distributions. The bottom panels show a comparison between the rolling percentiles of clusters at different ages. The outlines of the distributions (5th and 95th rolling-percentiles) are compared in the bottom-left panel, and their moving averages (50th rolling percentile) are compared in the bottom-right panel. Masses are derived as described in Appendix \ref{app:masses}. }
        \label{fig:observational_periodmass}
    \end{figure*}
    
    External photoevaporation of disks will also shorten the  star-disk-interaction (SDI) phase. During this phase, the magnetic interaction between stars and their accretion disks leads to an exchange of mass and angular momentum between the two \citep[\emph{e.g.,}][]{1979GhoshLamb,1991Koenigl,2021Ireland}. Stars lose the vast majority of their angular momentum during their first $\sim$10 Myr, and it is thought that the SDI is the primary driver of the stars' angular momentum evolution \citep{1995Bodenheimer,2004Mathieu,2014BouvierReview}. Physical models for the magnetic SDI show that the net result of this interaction can either spin up or spin down the star, driving the star's rotation toward an equilibrium spin rate \citep{1979GhoshLamb,2005MattAcc,2009ZanniFerreira,2013ZanniFerreira,2021Ireland}. 
    In this theoretical equilibrium state, the rotation rate of the star is dictated by a balance of torques, which is determined by physical parameters such as the accretion rate and stellar magnetic field strength.  The existence of an equilibrium spin rate is often simplified by an assumption, called "disk-locking" \citep{1993Edwards}, that the rotation rate of a star will remain constant for the duration of the accreting phase, \emph{i.e.}, for the lifetime of the disk \citep{1997Bouvier,2002Tinker,2004Rebull}. Therefore, by impacting the disk lifetime and duration of the SDI, the influence of the environment on the disk-dissipation timescales should also impact the host stars' angular momentum evolution.

    The importance of the SDI for removing angular momentum is supported by a growing number of observational studies in star-forming regions, which typically find a positive correlation between slow-rotation and the presence of a disk for solar-type stars ($M_*\sim 0.3-1.3$ M$_\odot$) along with statistically relevant differences in the distributions of observed spin rates for coeval stars with and without disks \citep[\emph{e.g.},][]{1992AttridgeHerbst,1993Edwards,2004Rebull,2006Rebull,2007CiezaBaliber,2008Irwin,2010Rodriguez-Ledesma,2017Roquette,2017Venuti,2018RebullUSco,2020Rebull}. The observational scenario is less settled for very-low-mass stars ($M_*<0.3$ M$_\odot$), and although most observational studies in this mass regime also point to a correlation between rotation and the presence of a disk \citep{2004Scholz,2015Scholz,2017Venuti,2019Moore,2018RebullUSco,2020Rebull}, this correlation is not observed in all circumstances \citep{2004Lamm,2010CodyHillenbrand}.

    Fig.~\ref{fig:observational_periodmass} illustrates the observational scenario typically used to constrain the spin evolution models of low mass stars by showing some of the most complete datasets to date. Period distributions are shown as a function of mass at 3 Myr, in NGC 2264 \citep{2005Lamm,2017Venuti}, at $\sim$8 Myr in Upper Sco \citep[USco,][]{2018RebullUSco}, at 13 Myr in h Per \citep{2013Moraux}, at 110 Myr in the Pleiades \citep{2016RebullPleiades}, at 710 Myr in the Praesepe \citep{2017RebullPraesepe}, and at 1 Gyr in NGC 6811 \citep{2019Curtis}. At the earliest stages of star formation, during the protostellar phase, we still have a poor understanding of what determines the stars' initial spin rates and what are the fundamental physical mechanisms responsible for their angular momentum changes \citep[\emph{e.g.},][]{2004WhiteHillenbrand,2005Covey}. However, once the stellar photosphere becomes visible, enabling the measurement of spin rates from the rotational modulation of spots at the stellar surface, 
    existing measurements of spin rates in star-forming regions reveal a broad distribution at all masses, with spin rates ranging from a fast rotation close to the break-up limit to slow rotators of about a dozen days \citep[\emph{e.g.},][]{2009IrwinBouvier,2014BouvierReview}.

    However, as broad as the initial spin rate distributions are, they soon become highly sub-structured. For example, most of the datasets in Fig.~\ref{fig:observational_periodmass} show some level of mass-dependency in their period-mass distributions. At the age of 3 Myr, the period-mass distribution of NGC 2264 hints at a dearth of slow rotators among the cluster's lowest mass stars ($\sim0.2\;\mathrm{M}_\odot$), with members under 0.4 M$_\odot$ rotating on average 2.2 d faster than members above this mass, and with no star rotating faster than one day among the highest mass ones. At the age of 8 Myr, the majority of USco members with masses under 0.4 M$_\odot$ are fast rotators, with a trend of faster rotation towards lower masses. This lack of slow rotators among the very-low-mass stars is observed in most clusters monitored by campaigns sensitive to these fainter stars.

    Once the disks are dissipated and the SDI phase is over, stars become free to spin up due to their PMS contraction until they reach the zero-age MS (ZAMS) and the ignition of hydrogen takes place. As lower-mass stars take longer to reach the ZAMS, they will spin up for longer periods of time, partially explaining the strong mass dependence observed at the Pleiades and Praesepe and visible in Fig. \ref{fig:observational_periodmass}. However, we still lack an explanation for the mass-dependence observed at the start of the spin-up phase.

    During the MS phase, the rotational evolution of stars becomes dominated by the spin-down due to mass loss via magnetised winds. At later MS ages  ($\gtrsim$1 Gyr), the rotation rates of solar-type stars are observed to converge toward a narrow range of rotation rates that decreases approximately as the square root of the age \citep{1972Skumanich}. This convergence happens in a mass-dependent way, \emph{i.e.,} more massive stars converge to the sequence at earlier ages than lower mass stars, and it suggests that the effect of the early stages on the rotational evolution is erased at late ages. The verification of this mass-period-age dependence has given rise to the concept of gyrochronology \citep{2003Barnes}, whereby the rotation rates of stars can be used to constrain ages, which is especially useful during the MS, when other stellar properties are less sensitive to age \citep{2016Barnes}. 
    
    When treated as an evolutionary sequence, the observed distributions of spin rates of different regions can be difficult to reconcile. For example,  \citet{2016Coker} demonstrated that the observed period-mass distribution at 13 Myr in h Per has an excess of fast-rotating stars that cannot be reproduced by using current spin evolution models to evolve the distribution of the ONC from 1 Myr to h Per's age.  Moreover, a recent study by \citet{2021Breimann} suggested that in order to be fitted by current models, present-day observed period-mass distributions of MS clusters require sub-structured initial conditions similar to the period-mass distribution of USco. The origins of both findings may be an environmental influence on the spin evolution of stars and, in parts, motivate this work. In particular, if the PMS environment does indeed influence stellar rotation on  $\gtrsim100$ Myr timescales, stellar rotation can offer a novel probe for the birth environment of stars and their planetary systems and help the investigations of growing claims in the literature for the environmental dependence of planet formation \citep[\emph{e.g.},][]{2016Brucalassi,2020Winter_Nature,2021Chevance,2021Longmore,2021Rodet}. 
    
    Working towards a better understanding of how the environment of star formation and early evolution can influence the spin evolution of low mass stars, we implement the results from a recent model for protoplanetary disk dispersal under the influence of external FUV-photoevaporation by \citeauthor{2020Winter} (2020, hereafter \citetalias{2020Winter}) into a spin-evolution model that treats the early-PMS phase under the disk-locking hypothesis. We apply this model to investigate the rotational evolution of low mass stars (0.1--1.3 M$_\odot$) from the early-PMS phase to the Sun's age and its dependence on the local FUV environment. 
    
    This paper is organised as follows. The implementation of our FUV-irradiated spin evolution model is described in Section \ref{sec:model}. In Section \ref{sec:results}, we present our results in both rotation-age (Section \ref{sec:basicmodels}) and rotation-mass (Section \ref{sec:results:mass_dependence}) parameter spaces. In Section \ref{sec:results_observations},  we compare our model results with the statistical properties of the observed period-mass distributions (shown in Fig. \ref{fig:observational_periodmass}). In Section \ref{sec:results_MassiveStars}, we add further context to our FUV-dependent spin evolution models by looking at the massive stars' influence on their local-FUV flux.  We discuss our model's limitations in Section \ref{sec:limitations} and debate the consequences of the influence of the environment on rotation in Section \ref{sec:consequences}. Finally, a summary and conclusions are presented in Section \ref{sec:SummaryConclusions}.
    
    \section{Model}\label{sec:model}
    
    To explore the role of environmentally induced variable disk dispersal timescales on the spin evolution of the central star, we require a model for the spin evolution that considers the presence of a disk during the early-PMS phase. Our base spin evolution model follows the general formulation from \citeauthor{2015Matt} (2015, hereafter \citetalias{2015Matt}). In this formulation, the spin evolution of the star is derived by using a forward-time-stepping Euler-method for solving the angular momentum differential equation for given initial conditions:
    
    \begin{equation}
     \frac{d\Omega}{dt}=\frac{T}{I}-\frac{\Omega}{I}\frac{dI}{dt}.\label{eq:spinevolution}
    \end{equation}
    
    \noindent Where $\Omega$ is the stellar angular velocity, $I$ is the stellar momentum of inertia, and $T$ represents the torques in action. 
    
    Our model is based on three assumptions. First, the stars are considered solid body rotators and their radius, $R_*$, their momentum of inertia, $I$, and its variation with time, $\frac{dI}{dt}$, are interpolated from a standard stellar evolutionary model as in \citetalias{2015Matt}, however, here we used the updated stellar evolutionary tracks of \cite{2015Baraffe}. 
     This solid-body approximation means that any angular momentum exchange with the exterior at the stellar surface is instantaneously communicated to the stellar interior. This approximation represents the limiting case of rapid internal angular momentum transport and neglects any possible effects from internal differential rotation.
    These limitations are discussed in Section \ref{sec:discussion:rigidbody}, along with alternative formulations.
    
     The second assumption is the hypothesis of the disk-locking scenario which is discussed and implemented in Section \ref{sec:diskmodel}. The third assumption is that once the disk-locking phase has finished, the only torque in action is the one from magnetised winds. The wind braking law adopted and its implementation are discussed in Section \ref{sec:wind}. Finally, we discussed the adopted initial conditions in Section \ref{sec:initial_conditions}.
    
    \subsection{Disk evolution}
    \label{sec:diskmodel}
    
    The disk-locking hypothesis has been widely adopted by previous models in the literature \citep[\emph{e.g.},][]{1997Bouvier,2013Gallet,2015GalletBouvier,2015VasconcelosBouvier,2017VasconcelosBouvier,2016Landin,2016Amard,2021Johnstone} and has the observational support discussed in the Section \ref{sec:introduction}. In this scenario,  as a consequence of the magnetic SDI, stars that are still accreting are exchanging angular momentum with their disks, and the net result of this interaction is a torque that counteracts the stellar contraction and stellar winds. Hence, while disk-locking is in action, \begin{equation}\label{eq:disk-locking}
        T_{\rm{SDI}}=\Omega\frac{dI}{dt} - T_W,
    \end{equation} and therefore $\frac{d\Omega}{dt}=0$. In our model, the star is kept with constant rotation for a time-scale $\tau_{\rm{D}}$ which will depend on details of the disk evolution process\footnote{Note that in this paper, we use the terms ``disk-locking duration'', ``disk dissipation timescale'' and ``disk lifetime'' interchangeably.}.  Alternative formulations for Equation (\ref{eq:disk-locking}) exist in the literature, including explicit formulations for the $T_\mathrm{SDI}$ and treatments for the SDI phase with $\frac{d\Omega}{dt}\neq0$. Those are discussed in Section \ref{sec:limitations:disklocking}, along with the limitations of the disk-locking hypothesis.
    
    In order to evaluate the duration of the SDI phase, we used results from the viscous-accretion disk dissipation models by \citetalias{2020Winter} which considers the evolution of an $\alpha$-disk that is subject to mass loss due to external photoevaporation at its outer radius and accretion of the material in the inner disk. \citetalias{2020Winter} considers a one-dimensional disk evolution as in \citet{2007Clarke}, and subsequently \citet{2018Winter}, with similar solutions to the disk evolution from \citet{1974Lynden-Bell}. In this model, the viscosity of the disk scales with its radius, with a scaling radius of 40AU, the mass of the star, $M_*$, and a viscous timescale $\tau_\mathrm{vis}\propto\frac{M_\mathrm{disk}}{\dot{M}_\mathrm{acc}}$ \citep[see also][]{2019Winter_CygOB2}. The disk has an initial mass of 0.1$M_*$, which is broadly consistent with the evidence that the initial mass of disks are related to the mass of the host star \citep[\emph{e.g.},][]{2000Natta,2006Natta,2009Fang,2011WilliamsCieza,2011Rigliaco,2013Andrews,2014Alcalca,2016Pascucci,2017Manara}. The disk evolves due to both accretion from the disk onto the star and external photoevaporation of material in the disk's outer edge, with the mass-loss rate due to external photoevaporation interpolated from the FRIED grid \citep[FUV Radiation Induced Evaporation of Discs,][]{2018Haworth}.  Other physical processes expected to give secondary contributions to the dissipation of disks are not included in \citetalias{2020Winter} but are discussed in section \ref{sec:discussion:diskdispersal}.
    
    \citetalias{2020Winter} evolved their disk model to derive FUV-induced disk destruction timescales, $\tau_{\rm{FUV}}$, calculated as a function of stellar mass, $M_*$, local FUV flux, $F_{\rm{FUV}}$, and viscosity timescale, $\tau_{\rm{vis}}$, with a disk considered dissipated when its mass becomes smaller than $10^{-5}$ M$_\odot$ and with $F_{\rm{FUV}}$ expressed in terms of Habing flux, G$_0=1.6\times10^{-3}$ erg cm$^{-2}$ s$^{-1}$ \citep{1968Habing}.  $F_{\rm{FUV}}$ values of 10, 100, 1000, 5000, and 10000 G$_0$ are considered. These are the values for which the FRIED grid is calculated, and cover the typical range of $F_{\rm{FUV}}$ expected in clusters \citep[10--10$^6$G$_0$, \emph{e.g.},][]{2008Fatuzzo,2018Winter}. The highest FUV flux considered was $F_{\rm{FUV}}=10^4$ G$_0$, with $\tau_{\rm{FUV}}$ saturating at this value. For larger fluxes than $10^4$ G$_0$, the temperature in the photodissociation of the disk becomes insensitive to the FUV irradiation and the mass flux and thermal winds will remain approximately constant \citep{1997Hollenbach,1998Johnstone}.
    A minimum disk dissipation timescale was therefore imposed imposed, with $\tau_{\rm{FUV}}$ saturating at $10^4$ G$_0$. A maximum timescale for disk dissipation of 10 Myr was also set, with the argument that if the disk survives external photoevaporation, it would be dissipated in this timescale due to internal processes. 
    
\begin{figure*}
        \centering
            \includegraphics[width=\textwidth]{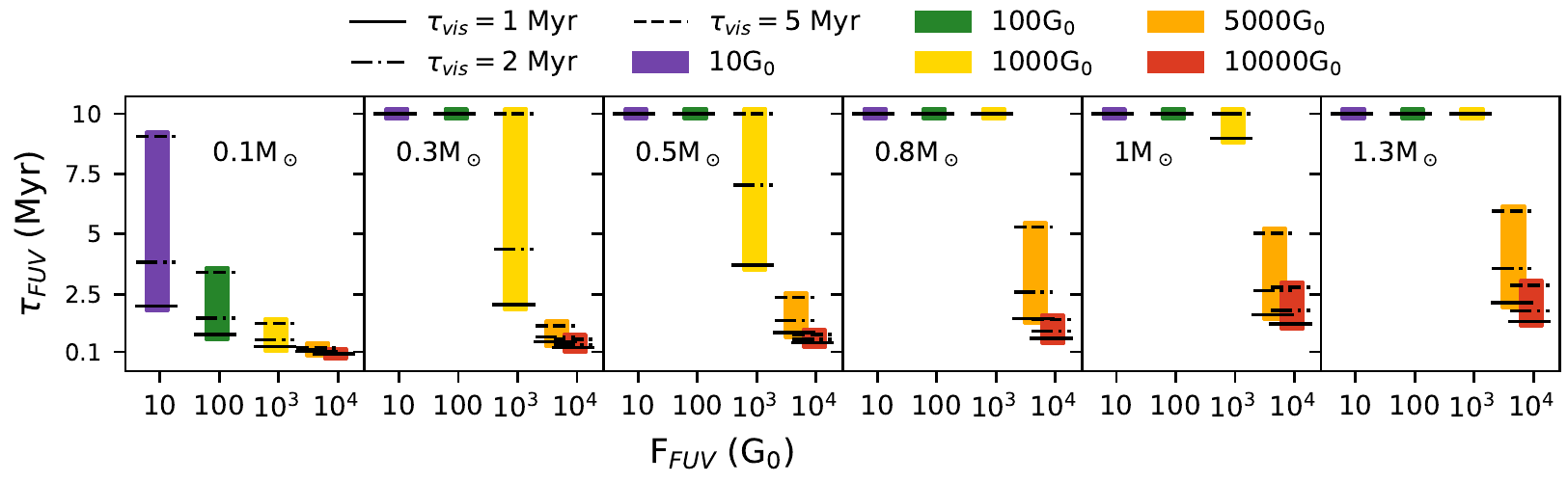}
        \caption{Time-scales for dissipation of viscous disks under the influence of external FUV-driven photoevaporation as a function of local FUV fluxes, which were used for constraining the disk-locking phase duration in our model. Results are shown for stellar masses, local FUV flux and viscous timescales at which models were calculated by \citetalias{2020Winter}: at 10, 100, 1000, 5000 and 10000 G$_0$, for stars with 0.1, 0.3, 0.5, 0.8, 1, and 1.3 M$_\odot$ masses, and disks with viscous timescale of 1, 2 or 5 Myr. Note that the bars are centred at the FUV-level they represent, however their width is set for visualisation purposes only.}
        \label{fig:tau_D}
    \end{figure*}

     Results for $\tau_{\rm{FUV}}$ are plotted as a function of $F_{\rm{FUV}}$ in Fig.~\ref{fig:tau_D} for six masses inside our mass range of interest, 0.1, 0.3, 0.5, 0.8, 1 and 1.3 M$_\odot$. The results are also shown for $\tau_{\rm{vis}}$ of 1, 2 and 5 Myr.  For further reference, the tabulations for $\tau_{\rm{FUV}}$ as a function of $F_{\rm{FUV}}$ presented in Fig.~\ref{fig:tau_D} are also provided in Table \ref{tab:DiskTabulations} in the Appendix. In the later development of our model we interpolated the appropriate $\tau_{\rm{FUV}}$ from Fig.~\ref{fig:tau_D} for intermediate masses and $F_{\rm{FUV}}$ values. 
    
     At a fixed viscous timescale, the disk-dissipation model adopted introduces a strong mass dependence on the timescales of disk dissipation. This mass dependence comes from the FUV-induced mass-loss rates from the FRIED grid, which at a fixed FUV-level are systematically higher for decreasing stellar masses. This happens because lower mass stars have shallower gravitational potentials, and consequently the material in their disks is less bound to the parent star \citep{2018Haworth}. For higher mass stars, the FUV-radiation is only efficient in photoevaporating the material in the disk down to certain radii that depend on the disk mass, disk viscosity, and on the FUV field strength. Hence, within the disk parameters covered by \citetalias{2020Winter}, stars with $M_*\geq0.3$ M$_\odot$ will be insensitive to external photoevaporation under low-FUV levels ($F_{\rm{FUV}}\leq100$ G$_0$, purple and green bars in Fig.~\ref{fig:tau_D}). For lower mass stars with $M_*<0.3$ M$_\odot$, material in the disk is externally photoevaporated more efficiently and down to smaller radii even under FUV-levels as low as 10 G$_0$. Consequently, for a 0.1 M$_\odot$ star, the maximum $\tau_{\rm{FUV}}$ is between 2 and 9 Myr, depending on $\tau_{\rm{vis}}$.
    
    Note that within the parameter space covered by the disk model adopted, some combinations of $F_{\rm{FUV}}$ and $\tau_{\rm{vis}}$ yield similar disk-dissipation timescales. For example, a 1 M$_\odot$ star has its disk dissipated in 2.8 Myr considering both $\tau_{\rm{vis}}=1$ Myr with a 4370 G$_0$ FUV level, and a  $\tau_{\rm{vis}}=5$ Myr with a 10$^4$ G$_0$ FUV level. Despite this, Fig.~\ref{fig:tau_D} shows that even though this degeneracy diminishes the strength of the mass-dependency of the disk-dissipation process for $M_*\geq0.3$ M$_\odot$ at the largest FUV levels, the overall mass dependence is still significant over the whole 0.1--1.3 M$_\odot$ mass range.

    \subsection{Wind torque}\label{sec:wind}
    
    The SDI ceases when the disk is dissipated after a timescale $\tau_{\rm{D}}\approx\tau_{\rm{FUV}}$, and after that, the only torque acting in the star is the torque from magnetised winds $T_\mathrm{W}$. The $T_\mathrm{W}$ adopted is the semi-empirically calibrated one of \citetalias{2015Matt}, which depends on the Rossby number of the stars, $R_0\equiv\frac{1}{\Omega_*\tau_\mathrm{cz}}$, and where $\tau_\mathrm{cz}$ is the convective turnover timescale:
    
    \begin{equation}
        T_W=
          \begin{cases}
    T_0\Big(\frac{\tau_{\mathrm{cz}}}{\tau_{\mathrm{cz}\odot}}\Big)^2\Big(\frac{\Omega_*}{\Omega_{\odot}}\Big)^{3}, &\quad \mathrm{if } \Omega_*\leq10\Omega_\odot\frac{\tau_{\mathrm{cz}\odot}}{\tau_\mathrm{cz}} \quad \mathrm{(unsaturated)} \\
    -100T_0\Big(\frac{\Omega_*}{\Omega_{\odot}}\Big), & \mathrm{otherwise.} \quad \quad \mathrm{(saturated)}
          \end{cases}
          \label{eq:torque}
    \end{equation}
    
    \noindent Where the torque scaling, $T_0=6.3\times10^{30}\mathrm{erg}\Big(\frac{R_*}{\mathrm{R}_\odot}\Big)^{3.1}\Big(\frac{M_*}{\mathrm{M}_\odot}\Big)^{0.5}$, is calibrated to the Sun. At each timestep, $\tau_\mathrm{cz}$ is computed using the prescription from \citet{2011CranmerSaar} based on the star's effective temperature.  The limitations of the prescription for $\tau_\mathrm{cz}$ are discussed in Section \ref{sec:limitations:taucz}. We refer to \citetalias{2015Matt} and references therein for a full discussion on the formulation of Equation (\ref{eq:torque}).  Because we updated the \citetalias{2015Matt} models to use the evolutionary tracks of \citet{2015Baraffe}, we had to also update the value for $\tau_{\mathrm{cz},\odot}$. This is required because the \citet{2015Baraffe} evolutionary tracks are not calibrated to the Sun, and therefore a $1$ M$_\odot$ star at the age of the Sun (4.56~Gyr) will have a slightly different effective temperature from the Sun, of 5716 K, and $\tau_{\mathrm{cz},\odot}=13.8$ d. 
    
    Finally, the rotational break-up limit is estimated as
    \begin{equation}\label{eq:breakup}
        \Omega_\mathrm{breakup}=\sqrt{\frac{GM_*}{(1.5R_*)^3}},
    \end{equation}
    \noindent following \citet[][Chapter 2]{2009MaederBook}. At spin rates close to this break-up limit, the material at the stellar equator becomes gravitationally unbound, which can induce augmented mass-loss rates and stellar wind geometry that differ from those in the wind-torque prescription adopted in this study. It is beyond the scope of this paper to explore how the torque would change close to break-up limit. Instead, we introduce a saturation condition in the spin-up phase, where stars are prevented from spinning faster than the break-up limit set by Equation (\ref{eq:breakup}). Whilst this is an over-simplification of the physics involved, it is effectively the same as assuming that stars rotating close to the break-up limit would have stronger wind torques that would efficiently brake them down, preventing them from rotating faster than the break-up limit. This assumption is required because, as presented in Fig.~\ref{fig:tau_D} and later discussed in this paper, under the highest FUV fluxes considered, the disk-locking duration can be significantly shortened even for the higher mass stars, leading to very long spin up phases that, otherwise, would result in stars exceeding the break-up limit.
    
    \subsection{Initial Conditions}
    \label{sec:initial_conditions}
    
    The lifetime of Class 0/I young stellar objects is estimated to be about 0.5 Myr \citep{2009Evans}. Hence, by starting our model after this phase, we can assume that the stars are past the phase of high and episodic accretion observed in Class I objects \citep[\emph{e.g.},][]{1977Herbig,2014Audard,2016Hartmann}, and therefore their disk evolution is well approached by the model described in Section \ref{sec:diskmodel}. Conveniently, the youngest age available for the whole 0.1--1.3 M$_\odot$ mass range in \citet{2015Baraffe}'s evolutionary tracks is 0.55 Myr, which is our starting point.
    
    At 0.55 Myr, we do not have direct observations for the period distributions. Instead, we based our choice of initial rotation rate on the earliest aged clusters observed like Taurus and $\rho$Oph \citep[$\lesssim\!3$ Myr;][]{2018RebullUSco,2020Rebull}, the ONC \citep[$\lesssim\!3$ Myr;][]{2009Rodriguez-Ledesma}, and NGC 2264 \citep[$\sim$3  Myr;][]{2017Venuti}, which exhibit broad period distributions from $\sim$16 d down to close to the break-up limit. While working under the disk-locking hypothesis, the range of rotation rates observed in disk-bearing stars during the PMS should reflect the range of initial rotation rates at the age of 0.55 Myr. For example, when considering the observed spin rates in $\rho$Oph, USco and Taurus, which are the closest young regions observed ($\lesssim$140 pc) and therefore have very complete samples of very-low mass stars, the majority of disk-bearing stars at all masses have rotation periods of about 1.6 d or larger. Motivated by that,  we considered two extreme values as initial conditions: A fastest initial spin rate of 1.6 d sets the minimum initial rotational period considered. This minimum value is compatible with the break-up limit at 0.55 Myr, which varies between 0.7 and 1.2 d depending on the stellar mass. For the maximum initial rotational period, we adopted the slowest rotation of 16 d, which reflects the maximum rotation rates typically observed in young regions.
    
    In Fig.~\ref{fig:tau_D}, some of the lower mass stars evolving under high FUV fluxes have their disk dissipated earlier than the age at which we start our model. For example, a 0.1 M$_\odot$ star evolving under 10,000 G$_0$ has its disk dissipated in 19,000 yr resulting in 0.53 Myr of spin up before the beginning of our spin evolution model at 0.55 Myr. This means that our initial conditions, which are based on the rotational properties of disk-bearing stars, might be too slow for such stars. However, these are also the models that reach the break-up limit during their spin-up phase. Therefore, enforcing faster initial conditions for those would only induce an earlier saturation of their rotation at the break-up limit.

    \section{Results}
    \label{sec:results}
    
    \subsection{FUV irradiated spin evolution}
    \label{sec:basicmodels}
    
    \begin{figure*}
        \centering
        \includegraphics[width=1.85\columnwidth]{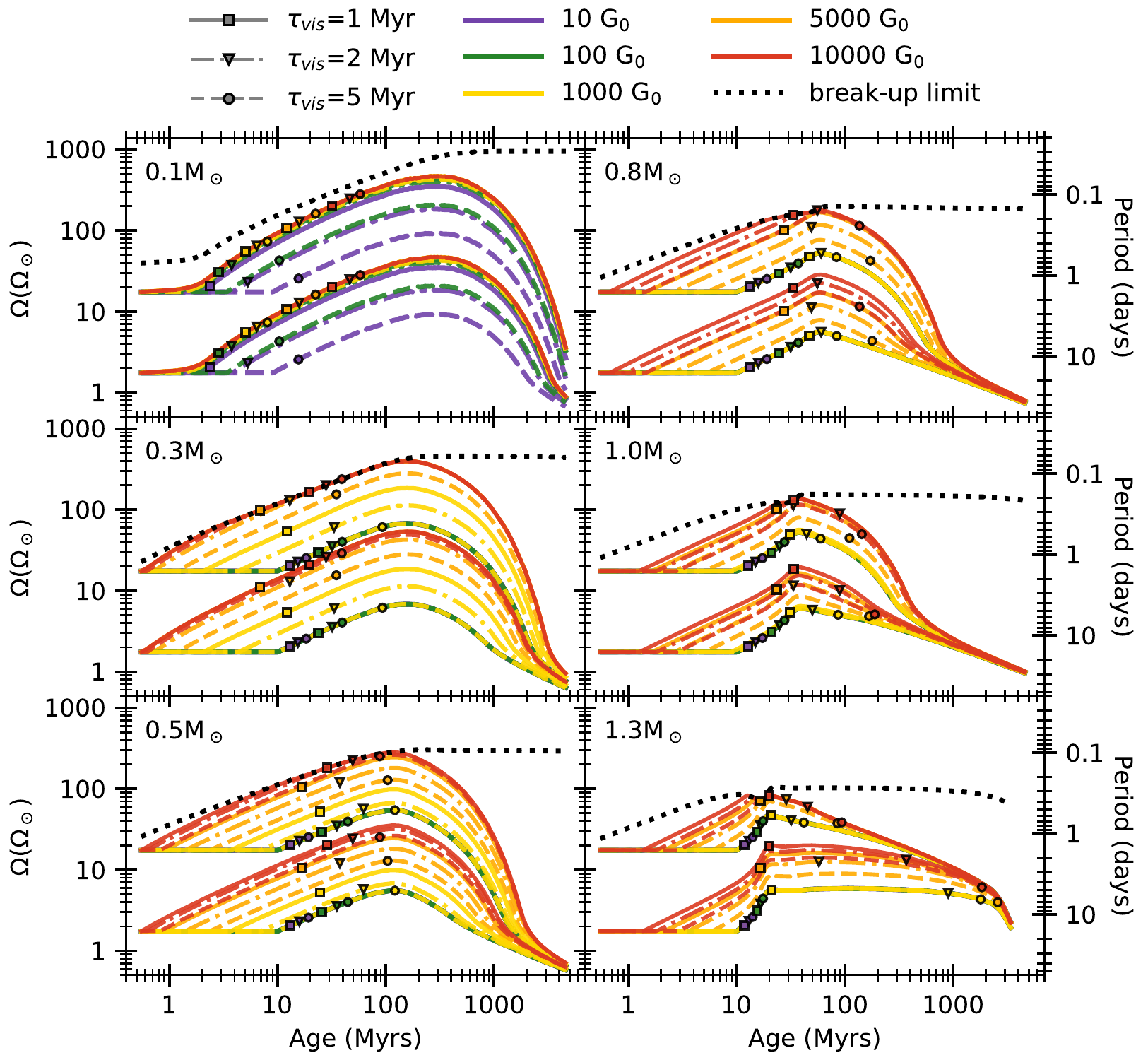}
        \caption{FUV-irradiated spin evolution models for stars with 0.1, 0.3, 0.5, 0.8, 1, and 1.3 M$_\odot$. The purple, green, yellow, orange and red lines present models for local FUV fluxes of 10, 100, 1000, 5000, and 10000 G$_0$ respectively. Models for initial rotation rates of 1.6 and 16 d are shown. Continuous, dashed, and dash-dotted lines show the models for $\tau_{\rm{vis}}$ of 1, 2 and 5 Myr, respectively. In order to facilitate the visualisation of models that fall on top of each other, a full squared, triangle or circle symbol is plotted at an arbitrary position for $\tau_{\rm{vis}}$ of 1, 2 and 5 Myr, respectively. The dashed black line shows the critical rotation limit for each mass.}
        \label{fig:spinmodel}
    \end{figure*}
    
    Fig.~\ref{fig:spinmodel} shows the results of the FUV irradiated spin-evolution model described in Section \ref{sec:model}, presented here with rotation rate as a function of stellar age. The models presented in Fig.~\ref{fig:spinmodel} are run using the two limit initial rotation rates discussed in Section \ref{sec:initial_conditions} (1.6 and 16 d), for each mass (0.1, 0.3, 0.5, 0.8, 1, and 1.3 M$_\odot$), local FUV flux (10, 100, 1000, 5000, and 10000 G$_0$), and $\tau_{\rm{vis}}$ (1, 2, and 5 Myr) for which the disk-dissipation timescales were estimated in Section \ref{sec:diskmodel}. 
    
    \subsubsection{Overall trends}
    
     The general trend seen in Fig.~\ref{fig:spinmodel} is that stars significantly affected by external photoevaporation are released from their disks earlier, having a reduced disk-locking duration. Consequently, these stars have a longer spin-up phase and reach the ZAMS rotating faster than stars with the same mass, viscous timescale, and initial condition but unaffected by external photoevaporation. 
     
    \subsubsection{Disk viscosity dependence}
    
    At a fixed mass and FUV flux, the disk-locking duration of stars having disks with shorter viscous timescales are more strongly reduced by external photoevaporation, as illustrated in Figs.~\ref{fig:tau_D} and \ref{fig:spinmodel}. This happens because the disk viscosity influences both the accretion rate and the expansion of the disk into a region where external photoevaporation is efficient \citep[see ][]{2019Winter_CygOB2}. Consequently, shorter viscous timescales generally result in higher accretion rates along with higher mass-loss rates due to external photoevaporation and systematically shorter disk-locking duration. Accordingly, at ages past the SDI, stars that had disks with lower viscous timescales tend to be faster rotators than stars that had larger viscous timescale. However this trend is only seen at specific combinations of stellar mass and FUV irradiation, as discussed in the following sections.
    
    \subsubsection{Solar-type stars}
    
    Considering stars with $M_*\geq0.3$ M$_\odot$, with the same viscous timescale, and under the same FUV flux, the effect of external photoevaporation will generally result in a shorter disk-locking duration for lower mass stars, and, consequently, at ages past the disk-locking phase, lower mass stars are systematically faster rotators. The only exception is the model for 1 M$_\odot$ stars with $\tau_\mathrm{vis}=$1 Myr, which has a disk dissipation timescale  $\sim\!1$ Myr shorter than the 0.8 M$_\odot$ star with the same viscous timescale. This variation is inherited from the FRIED grid \citep[see fig.~A3 in][]{2018Haworth}, which, as the disk evolves and its mass reduces to less than $10^{-4}$ M$_\odot$, gives higher mass-loss rates for the 1 M$_\odot$ star than the 0.8 M$_\odot$, resulting in a quicker disk-dissipation for the former. Consequently, for $\tau_\mathrm{vis}=$1 Myr models, 1 M$_\odot$ stars will reach the age of h Per rotating about 10$\%$ faster than 0.8 M$_\odot$ stars.
    
    Under low FUV fluxes of 10 and 100 G$_0$ (purple and green lines), all $M_*\geq0.3$ M$_\odot$ models are unaffected by external photoevaporation regardless of their viscous timescale. Under an intermediate FUV flux of 1,000 G$_0$ (yellow lines), models for 0.3 and 0.5 M$_\odot$ with $\tau_{\rm{vis}}=5$ Myr, and models for masses larger or equal 0.8 M$_\odot$ are also not significantly affected by external photoevaporation, but models for 0.3 and 0.5 M$_\odot$, and $\tau_{\rm{vis}}=1$ and 2 Myr have their disk-locking phase significantly shortened. All the stars are affected when under high FUV fluxes such as 5,000 or 10,000 G$_0$ (orange and red lines). 
    
    \subsubsection{Very-low-mass stars}
    
    Results for the 0.1 M$_\odot$ models show a distinct behaviour. During the early PMS, when the protostellar centre reaches a temperature of the order of $10^6$ K, the star is fuelled by deuterium burning through the reaction $^2$H$(p,\gamma)^3$He. The released nuclear energy is able to temporarily halt or at least slow down the contraction of the star. As it is sensitive to the central temperature, this phenomenon happens later for lower mass stars. In the case of the 0.1 M$_\odot$ star model, this stalling can last for more than a million years, during which the stellar radius barely changes, and so does its moment of inertia since the star is fully convective. For 0.1 M$_\odot$ stars evolving under 10 G$_0$, or 0.1 M$_\odot$ stars with  $\tau_{\rm{vis}}=5$ Myr disks evolving under 100 G$_0$, the deuterium-burning phase happens while the star is still locked to its disk. However, for all the remaining 0.1 M$_\odot$ models, the disk is dissipated between 19,000 yrs and 1.5 Myr, \emph{i.e.}, before the deuterium burning phase is over. Consequently, regardless of the disk being dissipated, 0.1 M$_\odot$ stars retain an approximately constant rotation rate until about 1.5 Myr of evolution, resulting in the overlapping 0.1 M$_\odot$ rotation models seen in Fig.~\ref{fig:spinmodel}. 
    
    \subsubsection{Saturation at the break-up limit}\label{sec:results:staturation_breakup}
    
    As a result of our model having a spin-up phase that saturates at the break-up limit (Equation (\ref{eq:breakup})), some of the highest FUV flux models will converge to a similar spin evolution once they approach their break-up limit. For stars with masses $M\geq0.8$ M$_\odot$, this happens a few million years before the ZAMS. 
    For the lower mass stars, when this convergence happens, it occurs significantly before the ZAMS. Note, however, that this convergence is an artefact of our model assumptions at extreme parameters and does not necessarily reflect a physical phenomenon (see discussion in Section \ref{sec:discussion:rigidbody}).
    
    \subsection{Spin evolution in period-mass space}
    \label{sec:results:mass_dependence}
    
    Next, it is relevant to discuss the stellar mass dependence of rotation in our model in the context of period-mass diagrams, as this is the preferred parameter space for observational studies. For that, we defined ``isogyrochrones'', which are tracks that cover the whole mass range modelled and show the spin rates for stars at a certain age, given a common initial condition, $\tau_{\rm{vis}}$ and local FUV flux. These tracks were estimated using a grid of masses between 0.1 and 1.3 M$_\odot$ sampled in steps of 0.02 M$_\odot$. For each FUV flux (10, 100, 1000, 5000 and 10000 G$_0$) and viscous timescale (1, 2, and 5 Myr), we defined a pair of fast and slow isogyrochrones with initial rotation of 1.6 and 16 d, which are aimed at describing the fast- and slow-rotation envelopes of the period-mass distributions. 
     
\begin{figure*}
        \centering
        \includegraphics[width=1.5\columnwidth]{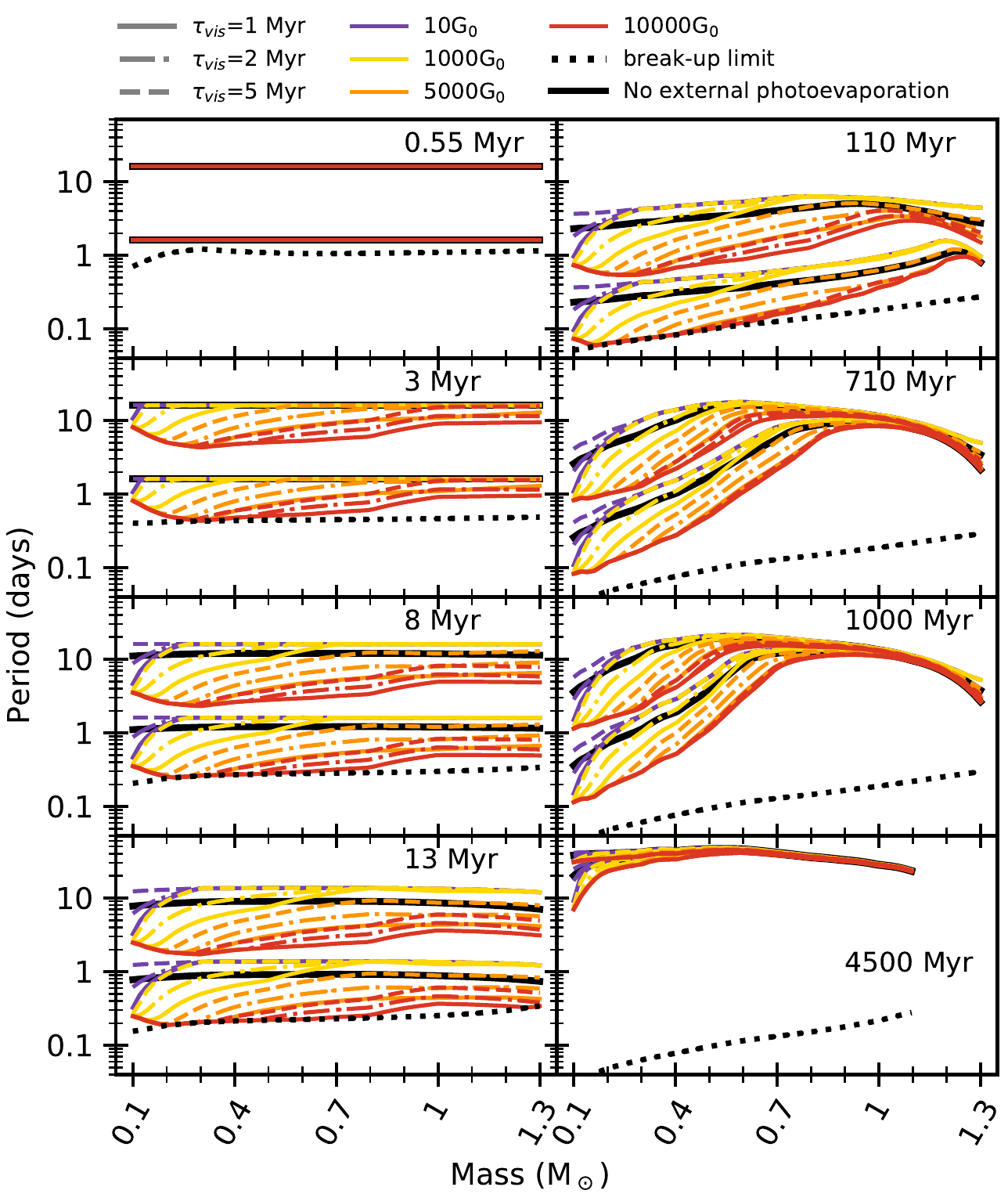}
        \caption{FUV-irradiated spin evolution model results in the period-mass space. Coloured lines show isogyrochrones for models at 4 different FUV fluxes: 10 G$_0$ (purple), 1,000 G$_0$ (yellow), 5,000 G$_0$ (orange), and 10,000 G$_0$ (red). At each FUV-level, models are presented for three values of viscous timescale: $\tau_{\rm{vis}}=$1 Myr (continuous lines), $\tau_{\rm{vis}}=$2 Myr (dash-dotted lines), and $\tau_{\rm{vis}}=$5 Myr (dashed lines).
        Each pair of isogyrochrones are for initial conditions with slow-rotation of 16 d (top lines) and fast-rotation of 1.6 d (bottom lines). For comparison, the black-continuous lines shows a model without the influence of external photoevaporation, with a fixed disk-locking duration of 5 Myr and independent of the properties of the stars. Models start at 0.55 Myr and are evolved until the age of the Sun (4.5 Gyrs), with snapshots shown at the ages of NGC 2264 (3 Myr), USco (8 Myr), h Per (13 Myr), Pleiades (110 Myr), Praesepe (710 Myr) and NGC 6811 (1 Gyr). The break-up limit at each age is shown as a dotted black line. Isogyrochrones for the FUV flux of 100 G$_0$ are omitted from this plot, as they are very similar to the 10 G$_0$ models.}
        \label{fig:spinmodel_tracks}
    \end{figure*}

    The results are shown in Fig.~\ref{fig:spinmodel_tracks} at the ages 0.55, 3, 8, 13, 110, 710, 1000, and 4500 Myr. While the initial age is simply the initial age of the evolutionary tracks of \citet{2015Baraffe}, discussed in Section \ref{sec:initial_conditions}, the subsequent ages shown are the approximate ages of the datasets in Fig.~\ref{fig:observational_periodmass} - NGC 2264, USco, h Per, Pleiades, Praesepe, NGC 6811 - and the Sun, respectively. For improving the visualisation, we omitted the 100 G$_0$ models from Fig.~\ref{fig:spinmodel_tracks}, as the models for 100 G$_0$ with $\tau_\mathrm{vis}$=1 and 2 Myr were very similar to the 10 G$_0$, $\tau_\mathrm{vis}$=1 Myr model and the 100 G$_0$, $\tau_\mathrm{vis}$=5 Myr model was very similar to the 10 G$_0$ $\tau_\mathrm{vis}$=2 Myr model. Models for 100 G$_0$ with $\tau_\mathrm{vis}$=1 Myr are later shown in Fig.~\ref{fig:observation_with_tracks}.
    
    \subsubsection{Control case: no external photoevaporation}
    
    For comparison, we also analysed a control case without external photoevaporation. For that, we considered a pair of isogyrochrones with initial rotation of 1.6 and 16 d and a fixed disk-locking duration of 5 Myr at all masses, which is equivalent to the \citetalias{2015Matt} model but revised under the disk-locking hypothesis. This control case is presented as bold black lines in Fig.~\ref{fig:spinmodel_tracks}.

    The isogyrochrones for the control case illustrate how, without external photoevaporation, the spin evolution only becomes mass-dependent close to the age of the Pleiades. This late mass dependence is also present in the spin evolution with external photoevaporation as it is inherent to \citetalias{2015Matt} wind torque (see Equation \ref{eq:torque}, Section \ref{sec:wind}), which has a dependence on stellar mass ($M_*$) and radius ($R_*$), and on the convective turnover timescale ($\tau_\mathrm{cz}$, which in turn depends on the effective temperature of the star). 
    However, the mass-dependency introduced by the \citetalias{2015Matt} torque only become evident at MS ages, when the spin evolution of stars is dominated by the wind-torque. Hence, the control case isogyrochrones in Fig.~\ref{fig:spinmodel_tracks} illustrates how without external photoevaporation the period distributions of young stars ($\lesssim 15$ Myr) are independent of stellar mass, which is inconsistent with the observed period-mass distributions at the age of NGC 2264 and USco (see also Figs.~\ref{fig:observational_periodmass} and ~\ref{fig:observation_with_tracks}). 
    
     \subsubsection{Trends for FUV-irradiated disks}
     
     In contrast to the control case, all FUV-irradiated isogyrochrones are mass-dependent since the early PMS phase, with the 10 G$_0$ model with $\tau_{\rm{vis}}=5$ Myr as the only exception. This dependency is independent of the initial rotation. Hence this is visible in both the fast and slow isogyrochrones, with a dearth of slowest rotators and an excess of faster rotators among the lowest mass stars. This feature is explained by the significant stellar mass dependency of the external photoevaporation of disks and its contribution to the disk-dissipation process, discussed in Section \ref{sec:diskmodel}, which under any fixed FUV flux results in an earlier start of the PMS spin-up phase for lower mass stars. Accordingly, Figs.~\ref{fig:tau_D} and \ref{fig:spinmodel_tracks} illustrate how the stellar mass dependency in the dissipation of FUV-irradiated disks translates into a mass-dependent PMS rotational evolution. A careful comparison between the two figures reveals a time-lag between a star being released from its disk and this resulting in a spin-up significant enough to be visible in the period-mass space. For example, 1 M$_\odot$ stars evolving under an FUV irradiation of 10,000 G$_0$ lose their disks in 1.3, 1.8, and 2.8 Myr for $\tau_{\rm{vis}}$ of 1, 2 and 5 Myr, respectively. Accordingly, at 3 Myr, the stars with a $\tau_{\rm{vis}}$ of 1 and 2 Myr already had 1.7 and 1.2 Myr to spin up, respectively, and have reached spin rates of 9 and 11.5 d. Meanwhile, the star with $\tau_{\rm{vis}}$=5 Myr just recently started spinning up, having a spin rate of 15.3 d, which is indistinguishable from its 16 d initial rotation within the typical precision of rotational surveys. This time lag is probably the explanation for the existence of slow-rotating diskless stars in some of the youngest regions observed \citep[\emph{e.g.}, NGC 2264, ][]{2017Venuti}. 
     
    The local FUV irradiation also dictates how early this mass-dependence is present in the period-mass diagram. Under intermediate and high FUV fluxes, this mass-dependency is already present at ages as early as 3 Myr. The mass range in which this mass-dependency is present increases through the early-PMS, and it is maximum at a few million years after stars at all masses and FUV fluxes have lost their disks. The extent of this mass-dependency increases with FUV flux until intermediate levels. At low FUV fluxes, disks are only mildly influenced by external photoevaporation, while under the highest FUV fluxes, all disks are quickly destroyed. In these extremes, disks under similar FUV fluxes have broadly similar disk-locking duration, and the resultant differences in rotation rates are relatively small. However, under intermediate FUV fluxes, the variations of external photoevaporation mass-loss rates as a function of stellar mass are large, so the beginning of the spin-up phase varies strongly between stars of different masses. For example, considering the $\tau_\mathrm{vis}=$1 Myr models, the ratio between the spin rate of a 1 M$_\odot$ and a 0.2 M$_\odot$ star by the age of h Per (13 Myr) is 1.4, 4.4 and 1.9 under FUV-fluxes of 10, $1,\!000$, and $10,\!000$ G$_0$, respectively.
      
    Finally, the mass-dependency in our models remain visible in the period-mass space even past the ZAMS, and it only disappears later on the MS, when higher mass stars start converging to a single slow-rotating sequence. Notably, while having longer spin-up phases, high-FUV models also have longer spin-down phases, such that, at a fixed age, the minimum mass at which the models converged to the slow sequence depends on the FUV irradiation of the model.
    
    \subsection{Comparing models to observed period-mass distributions} \label{sec:results_observations}
    
    \begin{figure*}
        \centering
         \includegraphics[width=1.5\columnwidth]{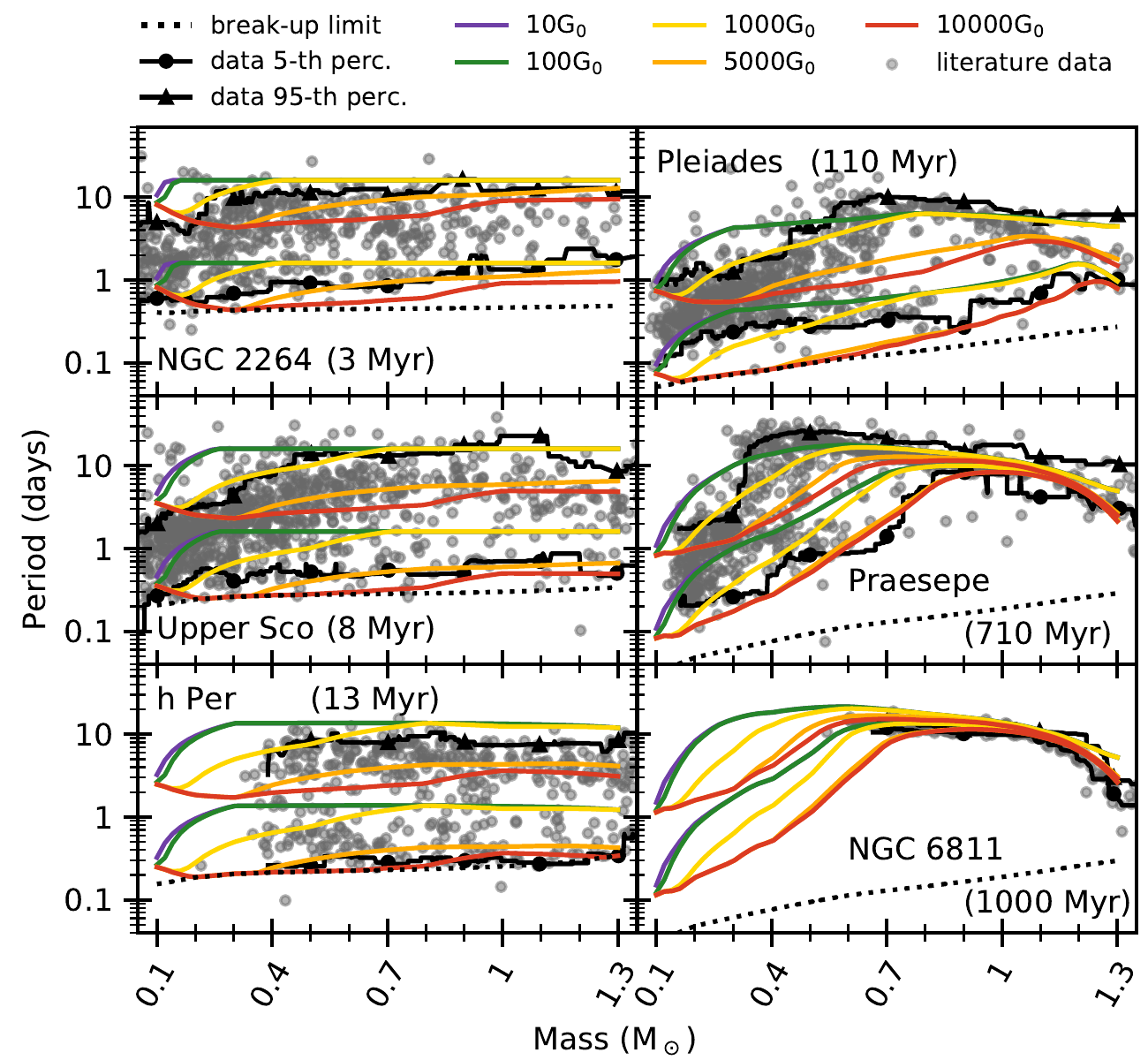}
        \caption{Observed period-mass distributions for NGC 2264, USco, h Per, Pleiades, Praesepe and NGC 6811, presented along with the isogyrochrones for models with $\tau_\mathrm{vis}=$1 Myr at different FUV-fluxes and at the age of each cluster,  with the 5th and 95th rolling percentiles of these distributions overplotted as in Fig.~\ref{fig:observational_periodmass}. Rolling-percentiles are calculated similarly to moving-averages, where the period-mass distributions are ordered as a function of ascending mass, with the percentile at a given mass estimated from a subset of observations around that mass, using subset sizes between 5-15$\%$ of the total number of stars in that observed dataset.}
        \label{fig:observation_with_tracks}
    \end{figure*}

      Fig.~\ref{fig:observation_with_tracks} shows the isogyrochrones for $\tau_\mathrm{vis}=$ 1 Myr, discussed in the previous Section, overplotted to the observed datasets. As in Fig.~\ref{fig:observational_periodmass}, we present observed datasets along with the rolling-percentiles of these period-mass distributions, with the 5th and 95th-percentile describing the distribution's fast and slow rotating envelopes and being comparable to the fast- and slow-isogyrochrones, respectively. We stress that this comparison is not aimed at reproducing the details of the substructures observed within these period-mass distributions, which would demand more detailed modelling of the initial distribution of spin rates and is beyond our scope.
      
    The general trend seen in Fig.~\ref{fig:observation_with_tracks} is that no model for single FUV flux can simultaneously explain both the 5th and 95th percentiles of observed clusters, indicating that to reproduce the statistical properties of observed period-mass distributions in clusters, prior knowledge of the distributions of FUV-fluxes within PMS clusters may be required. As the 5th percentiles of most clusters are better approached by higher-FUV fast-isogyrochrones, while the 95th percentiles are better approached by intermediate to low FUV slow isogyrochrones, a scenario emerges in which the fast-rotating population of clusters seem to be composed by stars that have evolved under high-FUV fluxes within a cluster, while the slow-rotating population has evolved under lower FUV fluxes. Accordingly, the outliers to the period-mass distributions can be explained as sources evolving under FUV-levels that diverge from the bulk of FUV-fluxes within a cluster. The scenario of a distribution of FUV-fluxes is further explored in Section \ref{sec:results_MassiveStars}.

    \subsubsection{Early-PMS ($\lesssim15$ Myr) regions}

    At 3 Myr, NGC 2264 period-mass distribution has a dearth of slow rotators among the stars with $M_*\leq0.3$ M$_\odot$, which is extended to $M_*\leq0.5$ M$_\odot$ by the age of USco. Fig.~\ref{fig:observation_with_tracks} shows that this behaviour can be reproduced by intermediate-FUV models, with the shape of the 1,000 G$_0$ slow-isogyrochrone approximately reflecting both NGC 2264 and USco slow rotators. 
    In particular, we note that the 95th rolling-percentile of USco reveals a period-mass dependence on the contrary sense for $M_*\gtrsim1$ M$_\odot$, with higher mass stars rotating faster than solar mass stars - previously noticed by \citet{2018RebullUSco} - which our models are unable to reproduce (see discussion in Section \ref{sec:discussion:diskdispersal}). At 13 Myr, the lack of observations for stars with $M_*\leq0.4$ M$_\odot$ in h Per hampers the comparison of its rotational distribution with the isogyrochrones for different FUV. Nevertheless, h Per slow rotators shows intermediate properties between the 1,000 G$_0$ and the 5,000 G$_0$ slow-isogyrochrones. 
    
    The shapes of the 5th rolling-percentile in the three younger regions are generally consistent with the shape of the highest FUV fast-isogyrochrones, and in the case of NGC 2264, our high-FUV models correctly reproduce the lack of stars rotating faster than 1 d among NGC 2264 stars with $M_*\gtrsim 0.9$ M$_\odot$. However, while at h Per's age, the 10,000 G$_0$ fast-isogyrochrone reflects the location of h Per's 5th rolling-percentile, in the case of NGC 2264 and USco, the initial condition of 1.6 d results in isogyrochrones too fast for describing the bulk of these region's fast rotators. Similarly, starting from the initial condition of 16 d allows reproducing the location of the slow-rotation envelope in USco, but results in a slow-isogyrochrone about 4 d too slow for NGC 2264.

     \subsubsection{MS regions ($\gtrsim 100$ Myr)}

     At later ages, the shapes of our isogyrochrones are less well matched to the observation's percentiles, which is due to a combination of missing physics in our model - specially regarding our solid-body assumption (see discussion in Section \ref{sec:discussion:rigidbody}) - and our simplified assumptions regarding the initial distributions of spin rates and single local FUV fluxes. Nevertheless, some of the general properties of the MS regions' period-mass distributions can still be approached by our models. 
     
     For example, at the age of the Pleiades (110 Myr), the mass dependence in the 95th percentile is well described by the 1,000G$_0$ isogyrochrone over the mass-range $M_*\lesssim0.5$ M$_\odot$. On the other hand, the 5th-percentile has intermediate properties between the intermediate and high-FUV models. By the age of Praesepe (710 Myr), the higher mass solar-type stars ($M_*\gtrsim 0.9$ M$_\odot$) are already converging to a narrow sequence of slow rotation. As the \citetalias{2015Matt} model was tuned to fit the Praesepe cluster, all isogyrochrones converge to reproduce the cluster's slow-sequence at Praesepe age. As discussed in Section \ref{sec:introduction}, this convergence happens in a mass-dependent way, and it means that the initial conditions - including the ones induced by the environment - are erased from the period-mass diagram. Nevertheless, as noted in Section \ref{sec:results:mass_dependence}, at a fixed age, the lowest mass at which some of the stars have started to converge to the slow-sequence will depend on both the initial conditions and on the local FUV fluxes. The faster the star's rotation is at the end of the PMS contraction phase, the longer it will take for the star to spin down under the torque of magnetised winds. As stars under the highest FUV levels will have longer spin-up phases, these will be the last ones to converge to the slow-sequence, and therefore the fingerprint of the initial conditions of stars submitted to higher local FUV flux during the PMS phase will be visible in the period-mass distributions for longer, which suggests that at a given mass and age, fast rotation rates for stars younger than a few Gyr could potentially work as a link between MS stars and their birth environment (see discussion in Section \ref{sec:exoplanets}).  Finally, by the age of NGC 6811, even though a significant mass range already converged to the slow-sequence, the converged isogyrochrones seem to rotate on average slower than the observed data, indicating that the models of \citetalias{2015Matt} need further modification (\emph{e.g.}, to the external torque or to relax the assumption of solid-body rotation).
    
    \subsection{The neighbourhood of massive stars}\label{sec:results_MassiveStars}
     
\begin{figure}
        \centering
        \includegraphics[width=\columnwidth]{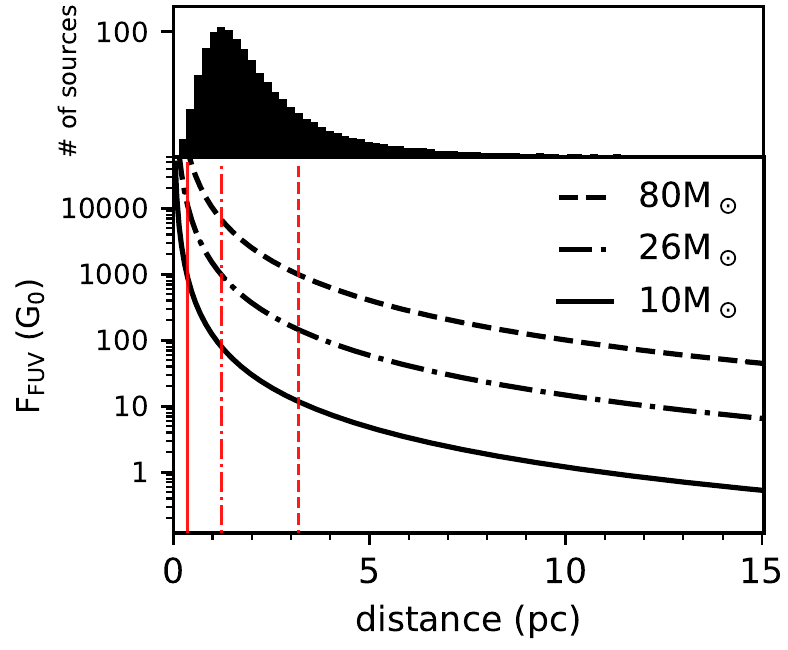}
            \caption{\emph{Bottom:} FUV flux variations in the proximity of three massive stars with 80 M$_\odot$ (dashed line), 26 M$_\odot$ (dash-dotted line), and 10 M$_\odot$ (continuous line). Local FUV fluxes are plotted as a function of distance from the massive stars in parsecs (black lines), while the vertical red lines show the distance at which each star contributes in 1,000 G$_0$ to their local flux. \emph{Top:} The histogram shows the distribution of 1392 sources in a Plummer Sphere with a scaling radius of 1.5 pc and central density of 100 sources per pc$^3$, which was used for simulating the evolution of low-mass stars in the neighbourhood of the massive stars in the plot.}
        \label{fig:OBstars}
    \end{figure}

     As discussed in the previous section, the mass-FUV-dependent disk-dissipation models introduced in our spin evolution code greatly contribute to explaining several features observed in the period-mass distributions of open clusters. However, Fig.~\ref{fig:observation_with_tracks} shows that none of the models for a single FUV-level could simultaneously explain both the 5th and 95th rolling-percentiles of the distributions and the outliers. This is not a surprise, as we do not expect the FUV environment of open clusters and star-forming regions to be described by a single FUV flux. At this point, it is useful to add context to the influence of massive stars on their local environment. For that, Fig.~\ref{fig:OBstars} portrays the influence of three massive stars with 80, 26 and 10 M$_\odot$ to their local FUV radiation field, where $F_{\rm{FUV}}=\frac{L_{\rm{FUV}}(M_*)}{4\pi r^2}$ is the local-FUV flux of a star with mass $M_*$ and FUV luminosity $L_{\rm{FUV}}(M_*)$, calculated at a distance $r$ from the star. We adopt stellar mass-dependent FUV luminosities as estimated as estimated by \citet{2003Parravano}, by integrating the stellar flux in the FUV bands using synthetic spectra and Padova models \citep{1994Bertelli}, and averaging the values through the MS lifetime of a massive star. Fig.~\ref{fig:OBstars} shows the contribution of massive stars to their FUV environment as a function of distance from the massive stars, and illustrates how this contribution is extremely mass-dependent. To facilitate the comparison, the vertical lines show the distance from the massive star where their local FUV flux will be increased by 1,000 G$_0$. This is at a distance of 3.20, 1.22, and 0.36 pc for stars with 80, 26, and 10 M$_\odot$, respectively.

    To examine the effect of massive stars on the rotational history of low mass stars in their surroundings, we considered the spin evolution of samples of low mass stars spatially distributed in a Plummer sphere \citep{1911Plummer} around each of the three massive stars in Fig.~\ref{fig:OBstars}. The distribution of sources in this Plummer sphere is set by:
    
    \begin{equation}\label{eq:plummer}
        \rho(r)=\frac{3N}{4\pi a^3}\big(1+\frac{r^2}{a^2}\big)^{-\frac{5}{2}},
    \end{equation}
    
    \noindent where $N$ is the total number of sources considered, $r$ is the distance to the centre of the sphere and $a$ is a scaling radius. The Plummer sphere considered has a scaling radius of $a=1.5$ pc, a central density of 100 stars per pc$^3$, and was truncated at a radius of 15 pc, which is the maximum distance considered in Fig.~\ref{fig:OBstars}. The distribution of sources in this Plummer sphere is shown in the histogram on the top of Fig.~\ref{fig:OBstars}. Local-FUV fluxes were estimated at each position based on the central massive star's FUV luminosity and the distance of each source to this central star. This distribution results in samples with 1392 sources evolving under FUV-fluxes in the range 45--558,682 G$_0$, 7--81,730 G$_0$, and 1--6,652 G$_0$ around the 80, 26 and 10M$_\odot$ stars, respectively. The median FUV fluxes were 2,695 G$_0$ for the 80M$_\odot$ star, 394 G$_0$ for the 26M$_\odot$, and 3 G$_0$ for the 10M$_\odot$.

\begin{figure}
        \centering
        \includegraphics[width=\columnwidth]{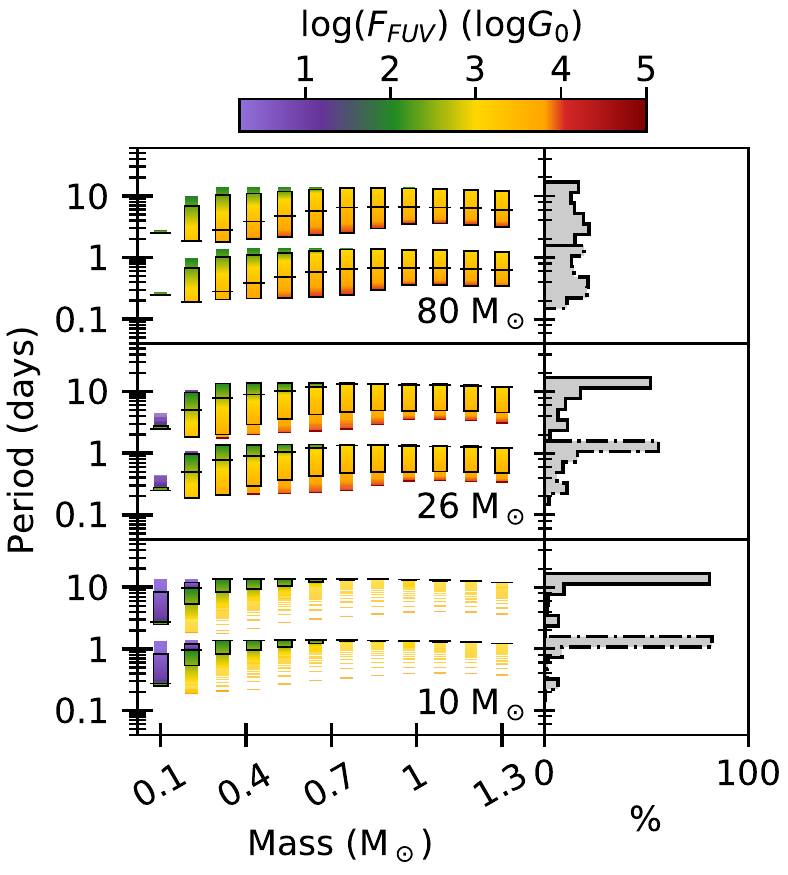}
        \caption{Period distributions at 13 Myr for stars evolving around the massive stars shown in Fig. 6, with 80 M$_\odot$ (top), 26 M$_\odot$ (middle), and 10 M$_\odot$ (bottom panels). Left panels show the period distributions of samples of 1065 equal-mass stars distributed in a Plummer Sphere around each massive star considering the fast and slow initial conditions of 1.6 and 16 d. Data points (`-' symbols) are coloured by the local FUV-flux at their position ($\log{(F_{\rm{FUV}})}$) around the massive star. The box-plots show the 95th (top horizontal line-segment), 50th (middle horizontal line-segment) and 5th percentiles (bottom horizontal line-segment) of the period distribution at each mass and initial condition. Right panels show histograms with the period distributions for the total of 16,704 stars simulated at all masses for the fast (continuous line) and slow (dash-dotted line) initial conditions.}
        \label{fig:PeriodMassOBstars}
    \end{figure}

    Next, considering stellar masses in the range 0.1--1.3 M$_\odot$ and using steps of 0.1 M$_\odot$, we placed same-mass stars at each of the 1392 positions in the Plummer sphere and modelled their spin evolution. The local FUV-flux at each position were used along with stellar masses and a fixed $\tau_{\rm{vis}}=1$ Myr to constrain their disk dissipation timescales, which were then used as an input to model the spin evolution of these stars. Models were run for fast and slow initial rotation rates of 1.6 or 16 d. Results of their spin evolution are presented in Fig.~\ref{fig:PeriodMassOBstars}, where we show the period distributions for each mass considered at the age of h Per (13 Myr).
    We point the reader to Section \ref{sec:limitations:cluster} for a discussion of the limitations of the model.
    
    Fig.~\ref{fig:PeriodMassOBstars} shows that the high-FUV environment of massive stars can 
    can skew the period distributions of low mass stars towards faster rotation. Within the context of the toy-model developed in this section, increasing the mass of the massive star at the centre of the Plummer sphere enhances the mass-dependence seen in the period distributions by extending the distribution's mass dependence over a wider mass range and resulting in a lower mass population rotating faster on average than around lower-mass massive stars.

    \section{Discussion }\label{sec:discussion}

    \subsection{Limitations in the model}\label{sec:limitations}
    
    \subsubsection{Disk-locking hypothesis}
    \label{sec:limitations:disklocking}
    
    In Section \ref{sec:diskmodel}, Equation (\ref{eq:disk-locking}) illustrated how we introduced the disk-locking hypothesis to our models, where the torques involved in the SDI, $T_\mathrm{SDI}$, counteract the wind-torque and PMS contraction, holding the stellar rotation constant during this interaction. While the same approach has been widely used in the literature, the disk-locking hypothesis is, in truth, an \emph{ad hoc} assumption meant to enforce the lack of spin-up in disk-bearing stars observed during the early-PMS phase while avoiding introducing the large uncertainties in the current theory for the SDI into the spin evolution models. In practice, Equation (\ref{eq:disk-locking}) assumes that as the star contracts and its momentum of inertia changes,  $T_\mathrm{SDI}$ will adjust itself in order to counteract this change and keep the rotation rate constant. Hence by setting the initial rotation rates as a free parameter in our model and looking at models with the initial rotation rates of 1.6 and 16 d, we are effectively looking at the extreme values of a range of possible $T_\mathrm{SDI}$. 
    
    A variety of physical mechanisms proposed in the literature could be part of $T_\mathrm{SDI}$. These are described by numerical models investigating the mechanisms of angular momentum transport during the SDI phase \citep[\emph{e.g.},][]{1993Collier,1994Yi,1996Armitage,2010Matt} as, for example, X-winds \citep{1994Shu}, accretion-powered stellar winds \citep{2012Matt}, and magnetospheric ejections \citep{2013ZanniFerreira,2019Gallet}. Alternatively, a study by \citet{2015Bouvier} also tried to account for the flux of angular momentum involved in the tidal and magnetic interactions between a 1 M$_\odot$ protostar and planets formed in its circumstellar disk. Yet, the majority of these models fail in predicting constant spin rates during the SDI phase, except under unrealistic physical conditions. 
    
    While none of these numerical models can currently simultaneously grasp all the physics behind the SDI, the main physical parameters responsible for this interaction are also still largely unknown. Consequently, current models for the SDI introduce too many free parameters in order to describe, for example, how accretion rates and magnetic fields vary with time. Additionally, these models are typically for stars at around one solar mass, and the parameter space covered by them is insufficient for a full description of the whole mass-range we investigated here. 
    
    Finally, the predicted evolution of spin rates during the SDI will likely be different from a constant rotation if full SDI torques are included in the models \citep[see][figs.~4 and 5]{2019Gallet}, but the qualitative results presented here should be robust,  \emph{i.e.}, if external photoevaporation removes a disk from a star at an earlier age, the effects of the SDI will end earlier, and the star will be more subject to spin-up from its own contraction, as predicted by our model.
    
    \subsubsection{Further mechanisms driving the dissipation of disks}\label{sec:discussion:diskdispersal}
    
    In the disk-dissipation model adopted in Section \ref{sec:diskmodel}, \citetalias{2020Winter} considered disks evolving due to the accretion process, along with the viscous spread of the material in the disk and FUV-induced external photoevaporation of the material on the outer edge of the disk. Fig.~\ref{fig:tau_D} showed that the disk-dissipation timescale variations with the local FUV-flux are mass-dependent at all FUV fluxes considered. This is translated into a mass dependent spin evolution, and explains the success of our model on reproducing the structures observed in the period-mass distributions of young clusters. 
    
    Yet, the disk-dissipation model adopted does not include all the physical mechanisms known to take part on the dissipation process of real disks. Other external drivers of disk photoevaporation, such as the extreme-UV radiation (EUV, $h\nu\geq13.5$ eV), can also contribute to the disks dissipation. However, we note that the mass-loss rates scale with the external radius of the disk to the power of $\frac{3}{2}$ for the EUV and roughly linearly for FUV \citep[\emph{e.g.},][]{1998Johnstone}, hence the EUV contribution is only relevant at the earlier disk evolution, and the FUV photons are expected to dominate over higher energy photons once the disk has been sufficiently truncated from the outer edge. As further explored by \citet[][see their fig. 12 and section 4.1]{2018Winter} a few exceptions exist, and  EUV should dominate over the FUV in extreme cases, either very close to high mass stars, where  $F_\mathrm{FUV}>10^4\;\rm{G}_0$ and the contribution of FUV to the disk dissipation reaches a plateau, or under very low $F_\mathrm{FUV}$ \citep[see also][]{1999StorzerHollenbach}. In the former case, the contribution of the EUV to the external photoevaporation would reduce even more the short $\tau_D$ for the $10^4$ G$_0$ case presented in Figure \ref{fig:tau_D}. In the latter case, while the EUV contribution would dominate, this contribution is still too small and should not significantly affect the derived $\tau_D$s. None of these cases should change the conclusions derived here.
    
    Dynamical encounters have also been posited as an alternative mechanism to induce premature dispersal of disks in dense environments \citep{2016Vincke}. However, random encounters are far less efficient at depleting disks than external photoevaporation \citep{2018Winter, 2019ConcharRamirez_b}. Encounters instead occur early during the dynamical decay of high order multiple systems to set initial conditions, rather than dispersal time-scales \citep{2018Bate}. Supernovae offer another potential avenue for externally induced disk destruction \citep{2017Close}, however empirical evidence for such destruction remains inconclusive \citep{2020Ansdell}. 
    
    Internal dispersal by mechanisms other than accretion also operate on protoplanetary disks and these may dominate the evolution at the end of the disk lifetime in the absence of external depletion \citep[\emph{e.g.},][]{2001Clarke}. Internal EUV \citep{2006Alexander}, X-ray \citep{2010Owen} and FUV irradiation \citep{2009Gorti,2009Gorti_B} may all play a role in late stage disk dispersal \citep[see][for a review]{2014Alexander}. In addition, some fraction of the disk mass must go into planets, the formation of which must therefore play a role in disk dispersal.
    
    The unknown contribution of these internal dispersal mechanisms to the disk-dissipation timescales may conflict with the assumption in our disk-dissipation model of a maximum $\tau_D$ of 10 Myr for disks around stars of all masses. In particular, a stellar mass-dependency of internal processes could help to explain the contrary-sense mass-dependency in the period-mass distribution of the Upper Sco region (Fig.~\ref{fig:observation_with_tracks} and Section \ref{sec:results_observations}), where 1 M$_\odot$ stars are on average slower rotators than 1.1--1.3 M$_\odot$ stars. Future studies considering the contribution of these internal processes may help improve spin-evolution models' capacity to fully reproduce the properties of observed datasets.  
    
    With the above considerations, the external FUV-driven winds should be interpreted as a process to systematically reduce the lifetimes of strongly irradiated disks in a statistical sense. Variations in disk evolution should introduce noise into this distribution, such that individual disks at single FUV flux may be depleted faster or slower by other influences. Nonetheless, observations indicate that gradients in disk lifetimes with FUV flux are still present \citep[e.g][]{2012Fang, 2016Guarcello}. The external irradiation of disks therefore remains a relevant factor in disk dispersal and therefore stellar rotation periods.

    \subsubsection{Rigid-body rotation and the rotational evolution of MS stars}\label{sec:discussion:rigidbody}
    
    In the \citet{2015Baraffe} evolutionary tracks, solar-type stars  ($M_*\geq0.3$ M$_\odot$) will eventually form a radiative core. Consequently, by assuming rigid-body rotation for the whole 0.1--1.3 M$_\odot$ range, we are treating these solar-type stars with their core and envelope rotating coupled to each other, with an instantaneous redistribution of angular momentum in the two regions. More robust treatments for the internal redistribution of angular momentum exist in the literature. For example, the widely used two-zone models \citep[\emph{e.g.},][]{1991MacGregorBrenner,1998Allain,2010DenissenkovB,2010Denissenkov,2011Spada,2013Gallet,2015GalletBouvier, 2015LanzafameSpada,2020SpadaLanzafame} consider the internal structure of solar-type stars with a radiative core and a convective zone, which both rotate as solid-bodies at all ages, but can assume distinct spin rates. Inside this framework, the core and envelope decouple shortly after the formation of the radiative core and, as the star evolves through the PMS, angular momentum gets hidden in a faster-rotating core while the slower rotating convective envelope is kept under the break-up limit \citep[see][]{2013Gallet,2015GalletBouvier}. Two-zone models are typically parametrised by a core-envelope coupling timescale, $\tau_\mathrm{ce}$, which indicates the rate of angular momentum exchange between the core and the envelope. Alternatively, a number of authors attempted to account for a full differential rotation in the radiative region 
    \citep[\emph{e.g.},][]{1990Pinsonneault,2010Denissenkov,2016SomersPinsonneault,2019Eggenberger,2019Amard,2021Gossage,2021Dumont}. However, we identified that every internal angular momentum transport description currently available in the literature include some initial condition or calibration constant derived while assuming that the environmental influence on rotation is negligible. 
    
    For example, most models include parameters, such as the $\tau_\mathrm{ce}$ in the two-zone models, derived by tuning or fitting spin evolution models to observed clusters. In this process, distributions of spin rates in clusters at different ages are assembled as an evolutionary sequence, which assumes that the environmental influence on rotation is negligible and that clusters with varied environments can be considered part of the same evolutionary sequence. \citet{2013Gallet,2015GalletBouvier}, for instance, had both $\tau_\mathrm{ce}$ and disk-locking duration as free parameters, which, along with the initial rotation and a calibration constant of the wind braking law, were simultaneously tuned to reproduce observations. Hence any disk dissipation timescales variation induced by the environment would go unnoticed behind the interplay between disk-dissipation timescale and $\tau_\mathrm{ce}$ used as free parameters. Alternatively, \citet{2015LanzafameSpada} derived a scaling relation between $\tau_\mathrm{ce}$ and stellar mass using the Monte Carlo Markov chain method to fit spin evolution models to observations \citep[see also][]{2020SpadaLanzafame}. However, their model assumed a fixed disk-locking duration of 5 Myr for all masses as an initial condition, which is inadequate to describe the rotation distributions of PMS clusters, as we have shown in Section \ref{sec:results:mass_dependence}. Finally, in order to avoid the uncertainties regarding the SDI phase, some authors start their spin-evolution model from the rotational distribution of h Per \citep[\emph{e.g.},][]{2016SomersPinsonneault, 2021Gossage}. However, we have shown in Section \ref{sec:results_observations} that h Per presents rotational properties compatible with a high-FUV environment, resulting in excess of fast rotating stars in h Per when compared to other clusters and it may not be an appropriate initial condition for MS clusters who had low or intermediate FUV environments during the PMS phase. 
    
    As the aim of this study was to probe the influence of the local-FUV environment on the spin evolution of stars, the model presented in Section \ref{sec:model} was designed while explicitly avoiding the inclusion of physical descriptions that are calibrated to reproduce distributions of spin rates in evolutionary sequences composed by clusters, such as the two-zone model for the internal structure of solar-type stars. By doing so, we were able to introduce the environmental influence on the SDI phase \emph{a priori} and explore its long-term impact on the rotation history of stars at different masses while avoiding biasing our model with physical prescriptions calibrated while ignoring environmental effects. 
    
    Avoiding a core-envelope decoupling prescription comes  at the expense of our models failing to reproduce some features visible in the period-mass distributions of MS clusters. One such feature is a dearth of stars observed at intermediate rotation periods between one and six days, forming a gap in the period-mass distributions of MS clusters, along with an enlarged $\Delta P$ between the fastest and slowest rotators at intermediate stellar masses. This gap is visible in the period-mass distributions of the Pleiades and Praesepe (Figs.~\ref{fig:observational_periodmass} and \ref{fig:observation_with_tracks}), and it suggests a more rapid spin-down just before convergence on the slow sequence \citep{2003Barnes}. A more rapid spin-down may be caused, for example, by a modification to the external torque \citep[\emph{e.g.},][]{2014Brown,2017Gondoin,2018Garraffo} or by a decoupling of the outer convection zone from the core \citep[\emph{i.e.}, by relaxing the assumption of solid-body rotation;][]{2013Gallet,2015GalletBouvier}.

    Another important issue is that at 1 Gyr, our models have spun down more than the observed stars in NGC 6811. The observational evidence for this stalled spin-down has been largely debated in recent years \citep{2018Agueros,2019Douglas,2019Curtis}, and a possible explanation for it is the interplay between core-envelope decoupling and the spin-down via magnetised winds. Such as, under the framework of the core-envelope theory, the resurfacing of angular momentum stored in the stellar core at later ages would compensate for the spin-down due to magnetised winds, temporarily stalling the stellar spin-down \citep{2020SpadaLanzafame}.

    Finally, considering treatments for the internal redistribution of angular momentum that differ from a rigid body would offer an alternative workaround for the assumption of spin-up saturation at the break-up limit introduced in Section \ref{sec:wind}. By allowing for angular momentum to be stored in a faster rotating core, the surface can be prevented from reaching the break-up limit at the end of the spin-up phase, which would naturally avoid the convergence of high-FUV models, discussed in Section \ref{sec:results:staturation_breakup}.
    
    \subsubsection{Convective Turnover timescale}\label{sec:limitations:taucz}
    
    A final caveat in the model is our choice of convective turnover timescale. By following \citetalias{2015Matt}, we adopted the prescriptions of \citet{2011CranmerSaar}, which provide $\tau_\mathrm{cz}$ as a function of effective temperature for the temperature range 3300--7000 K. Yet, we adopted their prescription even for stars down to 0.1 M$_\odot$, which have lower effective temperatures. \citet{2011CranmerSaar} provide comparisons between their estimations of $\tau_\mathrm{cz}$ and other literature derivations (see \citet{2011CranmerSaar} fig.~6). Their $\tau_\mathrm{cz}\sim$70 Myr for a $\sim$3000 K star is consistent with derivations by \citet{2009Reiners} for M-dwarfs, but it is 2--3 times lower than the derivations by \citet{2010BarnesKim}. The impact of an underestimated $\tau_\mathrm{cz}$ for the lowest mass stars can be discussed in the context of Equation (\ref{eq:torque}). An increase of 2--3 times in $\tau_\mathrm{cz}$ in our models would reduce the saturation limit by the same factor and increase the spin-down torque of unsaturated stars. In our models for 0.1 M$_\odot$ with a slow initial rotation of 16 d, stars spend their first $\sim$2 Myr in the unsaturated regime but become saturated as soon as they start to spin up, only returning to the unsaturated regime after $\sim$3 Gyr of spin evolution. At the earliest stages, the contribution of the wind-torque to the spin evolution of PMS stars is minimal. At the later stages, the 2--3 factor increase in $\tau_\mathrm{cz}$ would cause the models to become unsaturated at later ages than the 4.5 Gyr final age of our model. None of these cases should inflict any significant changes to our results or conclusions.
    
    \subsubsection{Distribution of low mass stars in the neighbourhood of massive stars}\label{sec:limitations:cluster}

    In Section \ref{sec:results_MassiveStars} we examined the influence of massive stars on the spin evolution of low-mass stars in their vicinity by looking at samples of equal-mass stars distributed in a Plummer sphere around single massive stars. In Fig.~\ref{fig:PeriodMassOBstars}, we presented snapshots at 13 Myr of the spin evolution of low-mass stars distributed around massive stars with 80, 26 and 10 M$_\odot$. These examples should be regarded as an exercise of application of our FUV-irradiated spin evolution models to generic populations rather than as attempts to precisely reproduce the FUV environments of real clusters. The latter would require more complex assumptions on the cluster geometry, population and their evolution, which are beyond the scope of this paper but will be the subject of a forthcoming paper. For the moment, we refer the reader to the growing literature exploring the FUV environments of clusters.
    
    \citet{2008Fatuzzo} have explored the dependence of the FUV-distributions in young embedded clusters on the size of the cluster (total number of members, $N$), the initial mass function (IMF) and the extinction. In the solar neighbourhood, clusters are small ($N\lesssim10^3$), and their IMFs are typically incomplete towards higher masses, with their FUV-distributions mainly dominated by the highest mass in the cluster. In this context, our examples with 10 and 26M$_\odot$ are a reasonable approximation for the expected scenarios in nearby clusters. In young massive clusters ($N\geq10^4$), larger populations of massive stars are present, and assumptions regarding the IMF are required to assess their FUV-distributions. In this regard, \citet{2008Fatuzzo} found that shallower IMFs result in clusters with larger populations of massive stars and larger outputs of FUV radiation. The existence of an 80M$_\odot$ star in a cluster population would thus accompany a number of other high-mass stars expected to also contribute to the cluster's FUV-distribution. Hence, for massive clusters, the rotational scenario is expected to be more complex than presented in Section \ref{sec:results_MassiveStars}, with broader distributions of FUV-fluxes resulting in wider distributions of $\tau_D$ and spin rates.
    
    The FUV-environments of real clusters has also been addressed by \citet{2018Winter}, who estimated FUV-fluxes as a function of local-number density in well-studied young clusters \citep[see also][]{2019Winter_CygOB2}. The example in Section \ref{sec:results_MassiveStars} considers a distribution of sources with a central density of 100 sources per pc$^3$, which only covers a small fraction of the number density vs FUV-flux expected in real clusters \citep[see][fig. 3]{2018Winter}. In this context, future studies should explore the rotational evolution of low mass stars in the context of this wider variety of cluster environments.
    
    We acknowledge the existence of other density distributions in the literature, such as the Elson-Fall-Freeman distribution \citep[EFF][]{1987Elson}, which are arguably a better fit for the distribution of stars in young clusters \citep[\emph{e.g.},][]{1987Elson,2003MackeyGilmoreA,2003MackeyGilmoreB}. The EFF distribution has a similar formulation to the Plummer sphere (Equation (\ref{eq:plummer})), but with an exponent of $-\frac{\gamma+1}{2}$ rather than $-\frac{5}{2}$. $\gamma$ describes how sharply the density of sources diminishes with $r$, and $\gamma\approx 1.3-4$ for young clusters \citep[\emph{c.f.},][table 1]{2018Winter}.  The EFF distribution is equivalent to a Plummer sphere when $\gamma=4$, and the example in Section \ref{sec:results_MassiveStars} thus runs on the upper limit for how centrally concentrated young clusters typically are. Lower values of $\gamma$ would result in shallower distributions of FUV and wider spreads in spin rates.
    
    Young clusters are embedded in parental gas and dust for a timescale of 1--3 Myr \citep{2007Allen}. The extinction caused by this material can reduce the FUV irradiation experienced by neighbouring stars \citep[\emph{e.g.},][]{2008Fatuzzo,2019AliHarries,2019Winter_CygOB2,2020Winter}, shielding disks from external photoevaporation for timescales of 1--2 Myr \citep[\emph{e.g.},][]{2007Clarke,2019Winter_CygOB2}. Our results, therefore, represent an upper limit for the fingerprints of the FUV-environments on rotation. Considering extinction should yield longer disk-dissipation timescales for disks affected by external photoevaporation,  thus changing both how early mass-dependency becomes evident in the period-mass distributions and the relative difference in $\tau_D$ at different masses. Nevertheless, this mass-dependency should be robust against the effects of extinction, as the differences in $\tau_D$ for extreme FUV regimes are larger than the timescale for which young clusters are still embedded. We also note that the distribution of parental material in clusters tends to be clumpy \citep[\emph{e.g.},][]{2003LadaLada}, with cluster members being unevenly affected by extinction, which could help to explain the spread of rotation rates observed in young clusters at a fixed mass.
    
    As high-mass stars evolve faster than low-mass stars,  the scenarios analysed in Section \ref{sec:results_MassiveStars} are only valid if disks affected by external photoevaporation are dissipated before the end of the high-mass stars lifetime. This question has been statistically addressed by \citet[][see their appendix A]{2020Winter}, who found that given the IMF of high-mass star-forming regions, even when their most massive star reaches the end of their lifetime, the region likely host at least another star responsible for an equivalent contribution to the local FUV radiation field, but with longer lifetime. In the case of the 80 M$_\odot$ star, while this star would have a lifetime of about 3.5 Myr, Figure \ref{fig:tau_D} has shown that under high FUV levels ($\gtrsim5000$ G$_0$), disks around low mass stars over a wide mass range would be dissipated in shorter timescales. Consequently, the results for stars within $\sim$1.5 pc from the massive stars (\emph{c.f.}, Fig. \ref{fig:OBstars}) would not be affected by considering the end of the massive star's life, while stars further away would have slightly longer $\tau_D$s.
    
    A final issue to be considered is that as clusters undergo dynamical evolution, the relative positions between cluster members change with time, as does the FUV irradiance at the position of each individual star. The external photoevaporation of disks in the context of the dynamical evolution of clusters has been approached in the context of different stellar densities by \citet{2019Concha-Ramirez,2021ConchaRamirezExternalPhotoevaporation}. Fluctuations in the local FUV fluxes at the position of individual stars induced by this dynamical evolution have also been illustrated by \citet[][see their fig~1.]{2021Parker}, who also analysed the evolution of disk-fractions for star-forming regions with varied initial conditions. While the impact of the initial density of sources on the average local FUV field is pointed to as the main factor governing disks evolution, \citeauthor{2021Parker} also demonstrated that small changes in the initial conditions of star-forming regions with identical initial densities could significantly change disks survival probabilities. Future studies should combine spin-evolution models with N-body simulations to address if the trends derived in this study are reliable within the context of dynamically evolving clusters.

    \subsection{Consequences of the environmental influence on rotation}\label{sec:consequences}

    \subsubsection{Environmentally dependent disk-locking duration}\label{sec:results:disklocking}
    
    An environmentally dependent disk-locking duration is the main novelty of our models. Nevertheless, the idea of a variable disk-locking duration is not new, and it has been explored by several studies presenting semi-empirical models for the spin evolution of low mass stars \citep{2015GalletBouvier,2017VasconcelosBouvier}. In particular, \citet{2013Gallet,2015GalletBouvier} used the disk lifetime as a free parameter in their spin evolution model. Starting from the 25th, 50th, and 90th percentiles (fast, median and slow rotators) of the rotational distribution of stars in the ONC ($\sim$1 Myr), they adjusted their disk-lifetime parameter to reproduce the properties of h Per rotational distribution at 13 Myr. They found that a variable disk-locking duration between 2 and 9 Myr was required to reproduce the ZAMS's broad rotational distributions. Their analysis focused on stars in the mass intervals 0.4-0.6 M$_\odot$, 0.7-0.9 M$_\odot$, and 0.9-1.1 M$_\odot$, and they found systematically longer disk-locking duration for the higher mass stars with median and slow rotation, but without significant difference with mass among the fast rotators.
    Notably, at a fixed mass range, \citeauthor{2015GalletBouvier} found shorter disk-locking timescales for faster initial spin rates. 
    As a possible explanation, they discuss how a correlation between initial rotation rate and disk-locking duration could be an outcome from the SDI operating at the embedded phase, with more massive disks yielding shorter initial rotation rates and living longer. Similarly, a series of studies by \citet{2015Tu} and \citet{2019Johnstone,2021Johnstone} derived and applied disk-dissipation timescales that scale with the initial spin rate of the stars, following the same trend of shorter disk lifetime for faster initial rotation. 
    
    In this paper, we give a step further by proposing that the variable disk-locking durations found by previous studies can be explained by the influence of the local FUV-field on the disk dissipation process. As discussed in Section \ref{sec:results_MassiveStars}, the presence of massive stars increases the FUV radiation field in their surroundings and introduces a distribution of local FUV-fluxes among neighbouring stars, shaped by the massive star luminosity and the relative positions between this massive star and its neighbours. In Figs.~\ref{fig:OBstars} and \ref{fig:PeriodMassOBstars}, we have shown that as a consequence of these variations in the local FUV fluxes, the period-mass distributions of low-mass stars in their neighbourhood gets skewed towards fast rotation, with the slow rotation envelope of the distributions dominated by stars under lower FUV fluxes, which at fixed mass had longer disk-locking duration, and the fast rotation envelope dominated by stars under higher FUV fluxes, which had shorter disk-locking duration. This result provides an alternative explanation to the correlations between disk-locking duration and initial rotation rate found by previous studies.

    All previous spin evolution models mentioned in this section include a two-zone description for the internal structure of solar-type stars, as described in Section \ref{sec:discussion:rigidbody}, with the core-envelope coupling timescale as a free parameter adjusted to fit the observed distributions of open clusters. Therefore the shorter disk-locking duration of fast-rotating stars obtained by these authors comes entangled with core-envelope coupling timescales that are shorter towards the faster rotators and longer towards lower mass stars. 
    We suggest that future studies should review these core-envelope coupling timescales to include the influence of the environment on the disk-dissipation timescales and to understand how this can impact other free parameters in the spin evolution models. This is important not only for a correct understanding of the core-envelope coupling phenomenology while considering the impact of the environment on the rotational history of stars, but also for improving our capacity of reproducing the rotational properties of MS clusters,
    
    \subsubsection{Implications for lithium depletion}
    
    A number of previous of studies have suggested that the amount of lithium depletion measured in solar-type stars is closely related to their rotational history \citep[\emph{e.g.}][]{2007Takeda,2008Bouvier}. In observations revealing a large dispersion of lithium abundances among stars in PMS and ZAMS clusters, fast-rotating stars are observed to be systematically less depleted in lithium than slowly rotating stars \citep[\emph{e.g.},][]{1987Butler,1988Balachandran,1993Soderblom,1997Jones,1998Jeffries,2016Bouvier,2016Messina,2018Bouvier}. These results have prompted the investigation of internal processes capable of explaining the trends observed, such as rotational mixing, internal magnetic fields and turbulence \citep[\emph{e.g.},][]{1990Pinsonneault,2010Denissenkov,2016SomersPinsonneault,2019Eggenberger,2021Dumont}. In particular, the fact that a lithium-rotation connection has been observed in clusters as young as NGC2264 \citep{2016Bouvier} suggests that a physical process already acting at such young ages must be behind the observed lithium-rotation connection. 
    
    The impact of the disk lifetime on the lithium depletion of solar-type stars has been previously addressed by \citet{2012Eggenberger}, who demonstrated that longer disk-locking durations increase the amount of differential rotation generated in the radiative zone during the disk-locking phase, resulting in more efficient rotational mixing and augmented lithium depletion during the PMS phase. Following this line, the environmentally induced variations on the disk-dissipation timescales explored in the present study offer a plausible explanation for the dispersion in lithium abundances observed in open clusters.

    An extra example supporting this hypothesis is the case of the lithium abundances in M67, which is a dense open cluster with a 3.5 Gyr age \citep{2018GaiaHRD}. As discussed by \citet{2016SomersPinsonneault}, M67 has a scatter in lithium abundances 1.4-4 times larger than younger clusters, depending on the effective temperature, that could not be reproduced by their spin evolution model calibrated by observations of younger clusters. \citeauthor{2016SomersPinsonneault} tested the hypothesis that M67 was richer in fast rotators than h Per, demonstrating that this intrinsically different rotational distribution at the age of h Per could explain the broad range of lithium abundances observed at 3.5 Gyrs in the M67. Indeed, a study by \citet{2005Hurley} used N-body simulations to model M67 from formation to its current age, finding that an initial cluster mass of $2\times10^4$ M$_\odot$ gave the best results. This suggests that the cluster had a rich massive star population during its early-PMS phase and was susceptible to large amounts of FUV-radiation, which inside the framework of our model results, could explain the distribution of spin rates skewed towards fast rotation at the age of h Per suggested by \citet{2016SomersPinsonneault}. 
    
    We suggest that future observational studies should explore the distribution of lithium abundances in solar-type stars in the vicinity of massive stars, which could help to establish a link between FUV irradiance, lithium depletion and the rotational history of low mass stars. Such a link could mean that the fast rotation of highly FUV irradiated PMS stars may leave imprints in older MS stars even after they have converged to the slow rotating sequence \citep[\emph{c.f.},][]{1997Pasquini}.
    
    \subsubsection{Open clusters at different ages may not be part of the same evolutionary sequence}
    
    Extending from the examples of the effect of massive stars in their FUV environment, discussed in Section \ref{sec:results_MassiveStars}, clusters with varied populations of high-mass stars will have distinct FUV environments and, consequently, different rotational histories. This result corroborates the suggestion of \citet{2016Coker} of a cosmic variance in the distribution of spin rates in different clusters. 
    For example, NGC 2264, which is the youngest cluster shown in Figs.~\ref{fig:observational_periodmass} and \ref{fig:observation_with_tracks}, has a total stellar mass of $\sim530$ M$_\odot$ \citep{2010Sung,2014Dib} and includes the the S Monoceros binary system as its highest mass star, which is composed of a 26 M$_\odot$ and a $\sim$9 M$_\odot$ component. In contrast, h Per currently has a cluster mass of $\sim1.6\times 10^4$ M$_\odot$ \citep{2010PortegiesZwart}. At 13 Myr, h Per has evolved beyond its first supernova explosion, having already lost its population with $M\gtrsim10$ M$_\odot$. Nevertheless, the cluster's current mass function suggests the previous existence of a numerous and diverse massive-star population. A high-FUV environment induced by numerous massive stars during the early-PMS of h Per could explain the excess of fast-rotating stars in the cluster, which are likely related to the stars that were the most affected by the FUV field in the cluster's early ages. 
    
    The contrasting properties of NGC 2264 and h Per highlights a bias existent in the literature of spin evolution of low mass stars. While most young stars are found in clusters and associations, only a small fraction of these will survive the first $\sim$10 Myr \citep{2003LadaLada}. Investigations of the physical processes leading to a cluster's disruption predict a dissolution timescale that scales with the initial mass of the cluster \citep[\emph{e.g.},][]{2006Lamers}. During the first 10 Myr, clusters and associations are easy targets for photometric monitoring surveys, resulting in observations of period-mass distributions of clusters with a wide variety of environments, but with a bias towards regions with reduced stellar density, since crowded regions cause confusion in photometric surveys. At later ages, the ensemble of clusters observable gets biased towards a reduced variety of initial cluster properties and higher initial cluster mass, like the case of the M67 cluster.
    
    From a theoretical perspective, in a follow-up study (Roquette et al., in prep), we will use realistic initial mass functions and source distributions to simulate the FUV environment of clusters with varied massive star content and explore their rotational properties.  From an observational perspective, we suggest that rotational surveys in young regions (3--10 Myr) rich in massive stars can help observationally establishing the role of high-FUV environments on the rotational history of low-mass stars. Furthermore, since the FUV-irradiance fingerprints on rotation are more robust for stars $M_\lesssim0.7$ M$\odot$, we highlight the importance of designing rotational surveys complete down to very-low mass stars. Finally, measuring the rotation rates for stars with $M_*\leq 0.4$ M$_\odot$ in h Per could significantly contribute to the field.
    
    \subsubsection{Implications and applications for exoplanet populations}\label{sec:exoplanets}
    
    Along with the stellar mass, rotation rate is empirically known to impact the X-ray emission and magnetic activity of low mass stars \citep{2011Wright}. \citet{2015Tu} demonstrated that initially fast-rotating stars remain magnetically active for longer periods of time than slow rotating stars. The influence of rotation on the X-ray, EUV and Ly-$\alpha$ emission can also impact the atmospheres of planets \citep[\emph{e.g.},][]{2015Johnstone,2021Johnstone}. For example, \citet{2020Johnstone} recently showed that the longer magnetic activity of fast rotating stars can enhance the escape of water-vapour from the atmosphere of Earth-mass planets in the habitable zone. A correlation between high-FUV environments and faster stellar rotation could, therefore, have important impact on the atmosphere of Earth-mass exoplanets.

    The premature destruction of disks by external photoevaporation may reduce the mass available for planet formation and even prevent planets from forming \citep{2005Youdin,2017Johansen,2017Ormel,2018Haworth_planet}. More generally, if the rotational periods of stars are correlated with the time-scale of disk depletion by external mechanisms then this offers a unique window into the early stages of planet formation for mature planetary systems. In particular, a number of recent studies have claimed that stellar kinematics are correlated with exoplanet architectures \citep{2020Winter_Nature, 2021Dai}, that may relate to high density stellar birth environments depending on the dispersal of kinematic substructure \citep[\emph{e.g.}][]{2019Kamdar}, efficiency of dynamical heating for open clusters \citep{2021Tarricq}, or initial clustering of open clusters in phase space \citep{2021Coronado}. Enhanced hot Jupiter frequency, for example, may be the result of dynamical perturbation in clustered environments \citep{2016Shara,2016Brucalassi,2021Longmore,2021Rodet}. However, the role of tidal inspiral and stellar ages on this finding remains debated \citep{2021Adibekyan,2021Mustill, 2021Winter}. 
    
    Stellar rotation periods offer a window into environment versus age for the younger ($\lesssim$1~Gyr) stellar streams or open clusters that can be aged more accurately than single stars. Within such a population, slowly rotating stars may have retained a disk for longer during formation, which may then link environment with, for example, the mass distribution of the exoplanets \citep{2021Kruijssen}. Hence, in light of our findings, we suggest that exoplanet surveys of open clusters \citep[\emph{e.g.}][]{2012Quinn,2014Quinn, 2016Malavolta,2020Takarada} with well quantified ages can be reconsidered in terms of the relative rotation periods of the surveyed stars. Such an exercise is particularly timely as \textit{Gaia} uncovers more stellar streams  \citep{2021Kamdar} and the TESS mission uncovers the low mass planet population \citep{2015Ricker}. 
    
    \subsubsection{The mass-dependency of rotation observed PMS clusters may not be a good proxy of age}
    
    The observation of mass-dependency in the period-mass distributions of young clusters along with the varying mass range in which this dependency is observed has led some authors to suggest that the slope of slow-rotation envelope of the period-mass distributions for stars $M_*\leq0.5$ M$_\odot$ could be used as a proxy of age \citep{2008Irwin, 2012HendersonStassun}. In light of the results presented in Sections \ref{sec:results_observations} and \ref{sec:results_MassiveStars}, we highlight that while this mass dependency does in fact evolve with age, the primary factor shaping this dependency is instead the FUV environment.

    \section{Summary and Conclusions}
    \label{sec:SummaryConclusions}
    
    This study investigated the influence of the local FUV environment around massive stars on the spin evolution of low mass stars. We did so by updating the \citet{2015Matt} spin-evolution model to include a disk-locking phase in which the stars exchange angular momentum with their disk through the star-disk-interaction, keeping a constant rotation rate during this interaction. To constrain the duration of the disk-locking phase, we used the results of disk-dissipation models by \citet{2020Winter}, which give the evolution of viscous disks under the influence of external photoevaporation driven by the local FUV-radiation (Section \ref{sec:diskmodel}). Due mainly to their shallower gravitational potential, external photoevaporation disperses disks around very-low-mass stars more quickly than those around higher mass solar-type stars (Fig.~\ref{fig:tau_D}).
    
    We modelled the spin-evolution of stars evolving under five FUV fluxes of 10, 100, 1000, 5000, and 10000 G$_0$, considering disks with one-tenth of the stellar mass and viscous timescales of 1, 2 and 5 Myr. We presented results in both rotation-age (Section \ref{sec:basicmodels} and Fig.~\ref{fig:spinmodel}) and rotation-mass (Section \ref{sec:results:mass_dependence} and Fig.~\ref{fig:spinmodel_tracks}) parameter spaces. Our results demonstrate that the mass-dependency introduced by the \citet{2020Winter} disk-dissipation model translates into an intrinsically mass-dependent rotational evolution since the early-PMS phase (Fig.~\ref{fig:spinmodel_tracks}). The mass range affected by external photoevaporation increases with the local FUV flux. The variation of spin rate with mass is the most prominent at intermediate local FUV fluxes of  1,000 G$_0$. For higher FUV fluxes of 5,000 or 10,000 G$_0$, stars of all masses lose their disks in less than 2--3 Myr of PMS evolution. Hence, at a given age, rotation rates are less dependent on mass at these high local FUV fluxes, but the period-mass distributions are faster-rotating overall when compared to lower local FUV fluxes. Future studies searching for observational evidence of the influence of the environment on rotation must acknowledge that to be observable in the period-mass distributions, stars that just got released from their disks need a few million years to spin up enough to become distinguishable from the stars still locked to their disks in terms of their spin up history. 
    
    The FUV-environment fingerprints are imprinted in the period-mass distributions until past the early-PMS phase and until MS stars converged to the slow-rotating sequence. At a fixed mass, stars influenced by their high-FUV environments started their spin-up phase earlier and, by having had more time to spin up, reached the ZAMS as faster rotators. At later MS ages, these stars will also be the last ones of their mass range to converge to a slowly rotating sequence. Consequently, at a fixed age, the minimum mass at which the period-mass distribution converges to a slowly rotating sequence is larger for stars with higher FUV levels during their early-PMS phase. This feature is relevant to the field of exoplanets, as it brings up rotation rates as an observable link to the planet formation conditions for MS stars of order Gyr ages. 
    
    By comparing our models to the 5th and 95th rolling-percentiles of observed period-mass distributions in real clusters (Section \ref{sec:results_observations} and Fig.~\ref{fig:observation_with_tracks}), we verified that no single FUV model could simultaneously describe both populations of fast and slow rotators. Nevertheless, intermediate-FUV models can be claimed to explain the mass-dependent slow-rotation envelope of the period distributions of USco and NGC 2264. Moreover, the excess of fast-rotating stars at 13 Myr in h Per can be explained by a high FUV environment during the cluster's early-PMS evolution. 
    
    Finally, we used samples of low-mass stars distributed in a Plummer sphere around massive stars to explore the rotational evolution in their neighbourhood. We demonstrated how the presence of massive stars is responsible for skewing the period distributions towards fast rotation while introducing augmented mass-dependency in the period-mass distributions around higher mass massive stars. Observational studies exploring the distribution of rotation rates in massive star-forming regions can help to confirm our results, especially if designed to include observations for very-low mass stars down to 0.1 M$_\odot$, which are the most affected by the FUV-environment.
    
    Our results provide a cautionary tale to previously published rotational evolution models with physical parameters calibrated by assembling observations of spin rates for stars in different clusters as an evolutionary sequence. In particular, we suggest that models exploring the influence of rotation on the internal mixing of stars should further examine the PMS environment's potential to induce cosmic variance in cluster's properties, such as lithium abundances. 
    
    \section*{Acknowledgements}

    We would like to thank Victor See, Mario Guarcello, Steven Rieder and 
    Guilherme Maluf for useful discussions during the development of this work. We also thank the anonymous referee for their constructive suggestions that helped to improve this manuscript. JR, SS, SPM, LA acknowledge funding from the European Research Council (ERC) under the European Union's Horizon 2020 research and innovation programme (grant agreement No 682393 AWESoMeStars). AJW acknowledges funding from an Alexander von Humboldt Stiftung Postdoctoral Research Fellowship.
    
    This research has made use of of the SIMBAD database \citep{Simbad}, and of the VizieR catalogue access tool \citep{VizieR}, both operated at CDS, Strasbourg, France, and of the 
    NASA's Astrophysics Data System Bibliographic Services.
    
    \emph{Software:} \texttt{astropy} \citep{astropy2013,astropy2018}, \texttt{matplotlib} \citep{Hunter2007}, \texttt{numpy} \citep{Harris2020}, \texttt{scipy} \citep{Virtanen2020}, \texttt{TOPCAT} \citep{Taylor2005}
    \section*{Data Availability}
    
    All the data and Python codes developed as part of this study are available as part of the package Far-ultraviolet Irradiated Rotational Evolution model for low mass stars (FIREstars) that can be accessible via \url{https://github.com/juliaroquette/FIREstars}. The package includes jupyter-notebooks with research notes on the project, computational tools for calculating spin-evolution models and isogyrochrones, along with the code used for producing each plot in the paper.
    
    
    
    \bibliographystyle{mnras}
    \bibliography{example} 

    
    
    
    \appendix
    
    \section{Observational Data}\label{app:masses}
    
    In this section we summarise the observational data used to compose Figs.~\ref{fig:observational_periodmass} and \ref{fig:observation_with_tracks}. The references for the rotational distributions used are presented along with the parameters for the clusters and associations adopted for estimating stellar masses.
    
    \subsection{NGC 2264}
    
    The NGC 2264 cluster is $\sim$3 Myr old \citep{2009Sung} and it is located at a distance $723^{+56}_{-49}$ pc \citep[Gaia DR2;][]{2018CantatGaudin}. NGC 2264 members had spin rates measured by multiple surveys, including ground- and space-based observations \citep[e.g.][]{1997Kearns,2004Makidon,2005Lamm,2013Affer,2016Sousa,2017Venuti}. We adopted primarily spin rates derived by \citet{2017Venuti} based on data from the CoRoT Mission. However, in order to populate the whole mass-range of interest, we complemented the \citeauthor{2017Venuti} dataset with spin rates measured by \citet{2005Lamm} using $I_c$ light-curves observed with the Wide Field Imager on the MPG/ESO 2.2 m telescope. The two surveys altogether provide spin rates for 522 stars in the mass range 0.1-1.3 M$_\odot$, 58$\%$ of these had individual extinctions estimated by \citet{2014Venuti}. When individual extinctions were not available, we adopted the \citet{2002Rebull} estimation of $E(B-V)=0.146$ mag for cluster members. We derived masses primarily by comparing photometric observations to the isochrones of \citet{2015Baraffe}. Pan-STARRS1 \citep{2020Flewelling_PS1} i-band data was available for $\sim$78.5$\%$ of the sample and was the main photometric band used for deriving masses. For the remaining stars, 12.7$\%$ used the Pan-STARRS1 g- or r- band, 6$\%$ used 2MASS \citep{2006Skrutskie_2MASS} J-band for disk-bearing stars, or Ks-band for diskless stars, and when neither Pan-STARRS1 or 2MASS data was available (2.6$\%$), we converted effective temperature estimations provided by \citet{2014Venuti} into masses.

    \subsection{Upper Sco}
    
    Spin rates for member of the Upper Sco association were obtained by \citet{2018RebullUSco} as part of the NASA K2 mission. Ages in the range 5-10 Myr have been proposed in the literature \citep[\emph{e.g.}][]{2002Preibisch,2016Feiden}, but we followed \citet{2018RebullUSco} and adopted an age of $\sim$8 Myr for USco members. A Gaia DR2 distance of 143 pc was adopted \citep{2019Damiani}. \citet{2018RebullUSco} also estimated individual extinctions for USco members using the 2MASS JHKs colour-colour diagram. When individual extinctions were not available, we adopted instead the association's mean reddening, $E(B-V)=0.245$ mag. \citet{2018RebullUSco} also verified the presence of a circumstellar disk in 21$\%$ of Upper Sco members. For diskless stars, we derived masses by comparing 2MASS Ks observations to the \citet{2015Baraffe}'s isochrones. For disk-bearing stars, in order to avoid the near-infrared excess in the longer wavelengths, we derived masses based on 2MASS J observations.
    
    \subsection{h Per}
    
    Spin rates for the open cluster h Per (NGC 869) were derived by \citet{2013Moraux} as part of the MONITOR project \citep{2007Aigrain_MONITOR}. We estimated masses for cluster members by comparing 2MASS Ks observation to the \citet{2015Baraffe}'s isochrones while adopting the cluster's mean reddening, $E(B-V)=0.54$ mag \citep{2010Currie,2013Moraux}, for all stars. We adopted an age of 13 Myr \citep{2008MayneNaylor,2013Moraux} for the cluster and the Gaia DR2 distance, $2336^{+712}_{-443}$ pc \citep[][]{2018CantatGaudin}.   
    
    \subsection{Pleiades}
    
    Spin rates for Pleaide's members were obtained by \citet{2016RebullPleiades} as part of the NASA K2 mission. We derived masses by comparing 2MASS Ks observation for cluster members to the \citet{2015Baraffe}'s isochrones, while adopting Gaia DR2 derivations of age (110 Myr), distance (136 pc), and reddening ($E(B-V)=0.045$ mag) \citep{2018GaiaHRD}. 
    
    \subsection{Praesepe}
    
    Spin rates for Praesepe's members were obtained by \citet{2017RebullPraesepe} as part of the NASA K2 mission. The age of Praesepe has been a matter of discussion in the literature. Isochrone fitting ages in the range $\sim$570-780 Myr have been previously obtain \citep[\emph{e.g.},][]{2011Delorme,2015BrandtHuang}, and a gyrochronology age of 670 Myr has been estimated by \citet{2019Douglas}. Here, we adopted the Gaia DR2 derivation of age for the cluster (710 Myr), which is somehow in the middle of other values found in the literature. We derived masses by comparing 2MASS Ks observation for cluster members to the \citet{2015Baraffe}'s isochrones,
    while adopting the aforementioned age along with the Gaia DR2 distance (186 pc), and reddening ($E(B-V)=0.027$ mag) \citep{2018GaiaHRD}.

    \subsection{NGC 6811}
    
    Spin rates for the open cluster NGC 6811 were obtained by \citet{2019Curtis} with data observed during the primary \emph{Kepler} mission. We derived masses by comparing Gaia DR2 G magnitudes for cluster members to the \citet{2015Baraffe}'s isochrones. We adopted an age of 1 Gyr \citep{2016Sandquist}, the Gaia DR2 distance $1113^{+139}_{-111}$ pc \citep{2018CantatGaudin}, and the cluster's mean extinction of $A_V=0.15$ mag \citep{2019Curtis}.

    \section{Tabulations for the disk-evolution timescale}\label{app:disktabulations}
    
    Table \ref{tab:DiskTabulations} presents tabulated results for the disk-dissipation models by \citet{2020Winter}, which is plotted in Fig.~\ref{fig:tau_D}. Disk-dissipation timescales ($\tau_\mathrm{FUV}$) are presented for stars with 0.1, 0.3, 0.5, 0.8, 1, and 1.3 M$_\odot$ evolving under FUV fluxes of 10, 100, 1000, 5000, and 10000 G$_0$ for disks with viscous timescales ($\tau_\mathrm{vis}$) of 1, 2, and 5 Myr.
    
    \begin{table}
    \centering
    \begin{tabular}{lrrr}
      $F_{\rm{FUV}}$ ( G$_0$) & \multicolumn{3}{c}{$\tau_{\rm{FUV}}$ (Myr)} \\
      \hline\hline
        & \multicolumn{1}{c|}{$\tau_{\rm{vis}}$=1 Myr}  &
      \multicolumn{1}{c|}{$\tau_{\rm{vis}}$=2 Myr}  &
      \multicolumn{1}{c|}{$\tau_{\rm{vis}}$=5 Myr} \\
     \cline{2-4}
        & \multicolumn{3}{c|}{0.1 M$_\odot$}\\
        \cline{2-4}
    10 & 2.001   &  3.832   &  9.050\\
    100  & 0.815 &  1.502   & 3.404\\
    1000 & 0.332 &  0.596   & 1.285\\
    5000 & 0.130 &  0.181   & 0.254\\
    10000 & 0.019&  0.019   & 0.020\\
    & \multicolumn{3}{c}{0.3 M$_\odot$} \\
    \cline{2-4}
    10 &  10.000   & 10.000 & 10.000\\
    100 & 10.000   & 10.000 & 10.000\\
    1000  & 2.068  & 4.360  & 10.000\\
    5000  & 0.522  & 0.722  & 1.186\\
    10000 & 0.271  & 0.391  & 0.623\\
    & \multicolumn{3}{c}{0.5 M$_\odot$} \\
    \cline{2-4}
    10    &  10.000  & 10.000 & 10.000\\
    100   &  10.000  & 10.000 & 10.000\\
    1000  &  3.699   & 7.032  & 10.000\\
    5000  &  0.902   & 1.408  & 2.375\\
    10000 &  0.481   & 0.628  & 0.818\\
    & \multicolumn{3}{c}{0.8 M$_\odot$} \\
    \cline{2-4}
    10      & 10.000& 10.000 & 10.000\\
    100     & 10.000& 10.000 & 10.000\\
    1000    & 10.000& 10.000 & 10.000\\
    5000    & 1.476 & 2.577  & 5.281\\
    10000   & 0.666 & 0.959  & 1.441\\
    & \multicolumn{3}{c}{1 M$_\odot$} \\
    \cline{2-4}
    10   & 10.000 & 10.000  & 10.000\\
    100  & 10.000 & 10.000  & 10.000\\
    1000 & 8.966  & 10.000  & 10.000\\
    5000 & 1.648  & 2.653   & 5.035\\
    10000& 1.255  & 1.817   & 2.779\\
    & \multicolumn{3}{c}{1.3 M$_\odot$} \\
    \cline{2-4}
    10   & 10.000 & 10.000 & 10.000\\
    100  & 10.000 & 10.000 & 10.000\\
    1000 & 10.000 & 10.000 & 10.000\\
    5000 & 2.149  & 3.567  & 5.959\\
    10000& 1.357  & 1.806  & 2.861\\
    \hline\end{tabular}
    \caption{\label{tab:DiskTabulations}Disk-dissipation timescales as a function of local FUV-flux and viscous timescale for different stellar masses.}
    \end{table}
    
    
    \bsp	
    \label{lastpage}
    \end{document}